\documentclass[preprint2]{aastex62}

\usepackage{mathtools}
\usepackage{rotating} 

\usepackage{booktabs} 

\usepackage{graphicx} 

\graphicspath{{./}{figures/}}
\usepackage{lineno}

\shorttitle{Constraining the NS Equation of State via SGRB X-ray Afterglows}
\shortauthors{}

\begin{document}

\title{Constraining the Neutron-Star Equation of State via Short Gamma-Ray Burst X-ray Afterglows}
\correspondingauthor{ Rahim Moradi; Yu Wang}

\author{R. Moradi}
\email{rmoradi@ihep.ac.cn}
\affil{State Key Laboratory of Particle Astrophysics, Institute of High Energy Physics, Chinese Academy of Sciences, Beijing 100049, China.}

\author{Y. Wang}
\email{yu.wang@icranet.org}
\affil{ICRA and Dipartimento di Fisica, Sapienza Universit\`a di Roma, P.le Aldo Moro 5, 00185 Rome, Italy.}
\affil{ICRANet, P.zza della Repubblica 10, 65122 Pescara, Italy.}
\affil{INAF -- Osservatorio Astronomico d'Abruzzo, Via M. Maggini snc, I-64100, Teramo, Italy.}

\author{F.~Rastegarnia}
\affil{ICRA and Dipartimento di Fisica, Sapienza Universit\`a di Roma, P.le Aldo Moro 5, 00185 Rome, Italy.}
\affil{ICRANet, P.zza della Repubblica 10, 65122 Pescara, Italy.}

\author{ E. S. Yorgancioglu}
\affil{State Key Laboratory of Particle Astrophysics, Institute of High Energy Physics, Chinese Academy of Sciences, Beijing 100049, China.}
\affil{University of Chinese Academy of Sciences, Chinese Academy of Sciences, Beijing 100049, China}

\author{Shu-Xu~Yi}
\affil{State Key Laboratory of Particle Astrophysics, Institute of High Energy Physics, Chinese Academy of Sciences, Beijing 100049, China.}

\author{B.~Eslam Panah}
\affil{Department of Theoretical Physics, Faculty of Science, University of Mazandaran, P.O. Box 47415-416, Babolsar, Iran.}

\author{S. N. Zhang}
\affil{State Key Laboratory of Particle Astrophysics, Institute of High Energy Physics, Chinese Academy of Sciences, Beijing 100049, China.}
\affil{University of Chinese Academy of Sciences, Chinese Academy of Sciences, Beijing 100049, China}

\begin{abstract}

Recent observations from NICER in X-rays and LIGO/Virgo in gravitational waves have provided critical constraints on the mass, radius, and tidal deformability of neutron stars, imposing stringent limits on the equation of state (EOS) and the behavior of ultradense matter. However, several key parameters influencing the EOS, such as the maximum mass of neutron stars, spin-down rates, and the potential role of exotic matter in their cores, remain subject of ongoing debate. Here we present a new approach to constraining the EOS by analyzing the X-ray afterglows of some short gamma-ray bursts, focusing on ``the internal plateau'' phase and its abrupt decay, which reflect the spin-down and possible collapse of a supra-massive neutron star into a black hole. By linking critical neutron star masses with black hole formation criteria and the observational data from Swift’s BAT and XRT instruments with compact object models,  we explore three representative EOSs that range from ``soft'' to ``stiff''. Our result supports a maximum mass for neutron stars of approximately 2.39 solar masses at the threshold of black hole formation. This conclusion holds under assumptions of magnetar-powered X-ray plateaus, constant radiative efficiency, isotropic emission, and full Kerr black hole energy extraction; deviations could influence the inferred results. Our results demonstrate the critical role of neutron star/black hole physics in probing dense nuclear matter and provide a novel framework for exploring extreme astrophysical environments.

\end{abstract}


\section{Introduction}\label{sec:1}

The neutron star equation of state (EOS) which relates pressure, density, and temperature at supranuclear densities, and constrains key global properties such as maximum mass, radius, and stability, is a powerful tool for probing the strong nuclear force under extreme conditions \citep{lattimer2001,2016ARA&A..54..401O,2024NatAs...8.1020M}. As specific examples, NICER observation  of PSR J0030+0451 yields $M=1.44^{+0.15}_{-0.14}\,M_\odot$ and $R=13.02^{+1.24}_{-1.06}\,$km \citep{2019ApJ...887L..24M}, while GW170817 constrains component masses $m_{1,2}\approx1.16$–$1.60\,M_\odot$ and tidal deformability $\tilde\Lambda\lesssim900$ \citep{2018PhRvL.120q2703A,2019PhRvX...9a1001A,2020PhLB..80335306L}. These multi‐messenger results have greatly narrowed the allowed EOS parameter space \citep{2018PhRvL.120q2703A}. Nonetheless, uncertainties still remain in the  density for quark matter \citep{2020NatPh..16..907A}, the maximum mass (e.g., $M_{\rm NS}\ge2.35\pm0.17\,M_\odot$ for PSR J0952–0607; \citealt{2022ApJ...934L..17R}), and centrifugal softening for spin rates $\nu\gtrsim700\,$Hz \citep{2024NatAs...8.1020M}. Resolving these issues requires refined dense matter modeling, and the development of novel theoretical and computational approaches.

Short gamma-ray bursts (SGRBs) present a unique observational avenue for studying the EOS of neutron stars. These energetic phenomena are widely associated with the merger of two neutron stars or a neutron star and a black hole \cite{eichler1989, narayan1992, 2004RvMP...76.1143P, 2014ARA&A..52...43B, 2018pgrb.book.....Z}. The merger dynamics produce a variety of remnants, ranging from prompt black holes to hypermassive or supramassive neutron stars (HMNS or SMNS) sustained by rotation and thermal pressure \citep{eichler1989,narayan1992,2018pgrb.book.....Z,rowlinson2013,2015ApJ...805...89L}. \textcolor{black}{One of the most intriguing observational features of some SGRBs is the presence of an internal plateau phase in their X-ray afterglows, lasting tens to hundreds of seconds. The magnetar spin-down interpretation of plateaus was first proposed for GRBs in general by \citet{2001ApJ...552L..35Z}, while their application to short GRBs was introduced by \citet{2006MNRAS.372L..19F}. The first observational evidence—together with the term \emph{internal plateau}—was reported by \citet{2007ApJ...665..599T}, with systematic searches by \citet{2007ApJ...670..565L} and \citet{2010MNRAS.402..705L}. This framework was later refined and expanded \citep[e.g.,][]{rowlinson2013, 2011A&A...526A.121D, lasky2014, 2019ApJ...872..114S, 2024A&A...692A..73G}. The sharp decline in X-ray luminosity following the plateau is consistent with the collapse of the SMNS into a black hole \citep{2013MNRAS.431.1745G, 2018MNRAS.475..266Z}; see Figure~\ref{fig:scheme}.} These observations not only constrain the physical properties of neutron stars, such as their maximum gravitational mass and spin frequency but also provide indirect evidence for the EOS. 



\textcolor{black}{While alternative explanations remain possible—including disk accretion mechanisms \citep{2011ApJ...734...35C}, fallback accretion, disk winds, jet evolution, early gravitational-wave losses \citep{lasky2014,2015PhRvL.115q1101M,2015ApJ...798L..36C,2015ApJ...802...95R,2019ApJ...880L..15M}, and viewing-angle effects \citep{2020MNRAS.492.2847B}—observational studies show that the magnetar spin-down model provides the most robust fits to SGRB plateaus \citep{2019ApJS..245....1T}.} For example,  analysis of SGRB X‑ray afterglows finds that they are very well modeled by magnetar spin-down energy injection, with only some outliers likely due to misclassification or alternative mechanisms \citep{2013MNRAS.431.1745G,2024A&A...692A..73G}. In contrast, models proposing energy injection from an accretion disk or the compatibility of the X-ray decline with the external shock model \citep{meszarosrees93, meszarosrees97, sari98, 2015PhRvL.115q1101M} offer potential explanations. However, these models face significant challenges. For example, short-lived massive disks struggle to produce the powerful, collimated jets observed in SGRBs \citep{2015PhRvL.115q1101M}. Additionally, the observed sharp X-ray decline is incompatible with the external shock model \citep{meszarosrees93, meszarosrees97, sari98}, which typically predicts a more gradual decay of the afterglow.

Moreover, the Dainotti relation—linking end-of-plateau luminosity and duration  in GRB afterglows \citep{2008MNRAS.391L..79D, 2010ApJ...722L.215D, 2011ApJ...730..135D, 2011MNRAS.418.2202D, 2013ApJ...774..157D,Dainotti:2022ked}—aligns  with magnetar spin-down models \citep{2019ApJS..245....1T, 2024ApJ...974...89W,Yorgancioglu2025Dainotti}. However, this interpretation depends critically on the assumptions of constant radiative efficiency and a neglected jet geometry \citep[see e.g.,][for more information]{Yorgancioglu2025Dainotti}. In conclusion, although uncertainties remain—such as the lack of definitive observations of fast radio bursts (FRBs) as the final signatures of SMNS in SGRBs \citep{2014A&A...562A.137F}—the SMNS collapse scenario continues to be the most compelling model for explaining the sharp decay in X-ray afterglows observed at the end of the internal plateau phase in SGRBs.


This work \textcolor{black}{builds upon previous studies that employ X-ray plateau durations to constrain the neutron star equation of state \citep{2014MNRAS.441.2433R,lasky2014,2015ApJ...805...89L,2022ApJ...939...51M,2024A&A...692A..73G}, but differs by implementing a magnetar multipolar electromagnetic field and by linking the critical neutron star mass to black hole formation criteria and applying black hole mass–energy relations, thereby providing a new framework for connecting electromagnetic observations of SGRBs with theoretical models of compact objects.} We use the internal plateau phase of SGRB light curves to trace the spin evolution of the SMNS and the post-collapse luminosity as a measure of the energy extracted from the newly formed Kerr black hole (KBH). Combining the critical neutron star mass formula with the KBH mass–energy relation \citep{1970PhRvL..25.1596C,1971PhRvD...4.3552C,1971PhRvL..26.1344H} allows us to determine the maximum gravitational mass, \( M_{\rm max} \), and spin parameter at collapse \citep{lasky2014,2015PhRvD..92b3007C}, offering a robust means of constraining the EOS.

The core assumptions of our model include: (1) the X‑ray plateau is powered by a magnetar multipolar electromagnetic field, (2) radiative efficiency remains constant and jet beaming is neglected (i.e., isotropic emission), and (3) complete extraction of KBH energy as electromagnetic radiation, neglecting gravitational‑wave emission. Deviations from these assumptions could alter the inferred Tolman–Oppenheimer–Volkoff (TOV) mass limit, $M_{\rm TOV}$.

The paper is structured as follows. In Section~\ref{sec:methods}, we present the methods for reproducing the bolometric light curve, the multipolar electromagnetic spin‑down formulation, and the luminosity‑fitting procedure. In Section~\ref{sec:results}, we introduce the theoretical framework used to constrain the KBH mass and spin at birth, which in turn places limits on the system’s equation of state. In Section~\ref{sec:conclusion}, we present concluding remarks, discuss caveats and strengths, and outline directions for future work.


\section{ Data and Methods}\label{sec:methods}

\subsection{Bolometric X‐ray Light Curves}\label{subsec:obs}



\textcolor{black}{In the study of GRBs, long-lived afterglow emission from the external shock in GRBs spans a broad range of wavelengths, including radio and optical bands \citep{2007A&A...469..379E,2012ApJ...746..156C,2018ApJS..234...26L}. In contrast, the internal plateau phase—originating at an inner radius—is characterized by emission that is typically self-absorbed at low frequencies and thus generally lacks detectable radio or optical counterparts. For example, the plateau in GRB 070110 shows no radio or optical emission \citep{2007ApJ...665..599T}, and GRB 090515 exhibited no optical afterglow during the X-ray plateau phase, with only a very faint optical detection ($r \approx 26.4$) appearing about 1.7 hours after the burst \citep{2010MNRAS.409..531R}. Moreover, since the internal plateau occurs near the prompt emission phase, its spectrum likely extends into the energy range covered by instruments such as Swift/BAT (10–350 keV) \citep{2018ApJS..235....4O} and Fermi/GBM (10 keV to 10 MeV) \citep{2014ApJS..211...12G,2020ApJ...893...46V}. }

\textcolor{black}{ Consequently a comprehensive estimate of the bolometric luminosity for the internal plateau requires extension of the energy band to this broad range. Therefore, the lack of low-frequency emission, the expected spectral properties at high energies, and the characteristic sharp decline in flux at the plateau’s end—which motivated the classification of internal plateaus by \citet{2007ApJ...665..599T}—support bolometric luminosity calculations over the 1–10000 keV range, where contributions below 1 keV are negligible \citep[e.g.,][]{rowlinson2013}. In previous work by \cite{2024ApJ...974...89W}, where Swift-XRT observations offered the most complete afterglow data, the X-ray afterglow luminosity in the 0.3–10 keV range was multiplied by a factor of five to approximate the bolometric luminosity. However, in this paper, { in order} to account for prompt emission data, we adopt the approach outlined by \cite{rowlinson2013}, which provides a more robust method for deriving bolometric luminosities.}

{\color{black}
For the analysis, the X-ray light curves within the 0.3--10 keV energy range were sourced from the automated data analysis tools provided by the UK Swift Science Data Centre \citep{Evans2007,Evans2009}, which offer preprocessed light curves for each individual SGRB. BAT (Burst Alert Telescope) light curves were generated using standard HEASOFT pipelines, with 3$\sigma$ significance bins employed. The spectra in the 15--150 keV range from BAT were fitted using \texttt{XSPEC} and subsequently extrapolated to estimate the flux in the 0.3--10 keV band. By combining this extrapolated flux with the count rates from the BAT spectra, each data point in the BAT light curve was rescaled to correspond to a 0.3--10 keV flux, utilizing a simple power-law spectral model.

These rescaled BAT light curves were then combined with the XRT light curves to produce a unified BAT--XRT light curve (see Figure~\ref{fig:LC}). To facilitate comparison with magnetar models, the observed light curves were converted to unabsorbed fluxes and subsequently transformed into rest-frame luminosity light curves in the 1--10000 keV range through a bolometric $k$-correction process \citep{2001AJ....121.2879B}.

We model the plateau-phase spectrum using the \cite{1993ApJ...413..281B} function:\\
\\
\[N(E) = \]
\[
\begin{cases}
A\left(\frac{E}{100\,\mathrm{keV}}\right)^\alpha e^{\left(-\frac{E}{E_0}\right)}, ~~~~~~~~~~~~~~~~~ E \le (\alpha - \beta) E_0, \\[0.3em]
A\left[\frac{(\alpha - \beta) E_0}{100\,\mathrm{keV}}\right]^{\alpha - \beta} e^{(\beta - \alpha)} \left(\frac{E}{100\,\mathrm{keV}}\right)^\beta, ~ E > (\alpha - \beta) E_0.
\end{cases}
\]

where $\alpha$ and $E_0 = \frac{E_{\rm peak}}{2 + \alpha}$ are obtained from the spectral fits. The bolometric $k$-correction factor is defined as

\[
k = \frac{\displaystyle \int_{E_1/(1+z)}^{E_2/(1+z)} E\,N(E)\,dE}{\displaystyle \int_{e_1/(1+z)}^{e_2/(1+z)} E\,N(E)\,dE},
\]
where the numerator corresponds to the rest-frame energy band of interest (1--10000 keV), and the denominator corresponds to the observed Swift/XRT band (0.3--10 keV). Both integrals are evaluated numerically for each burst using its specific spectral parameters $(\alpha, \beta, E_0)$.

Because Swift/XRT data typically constrain only $\alpha$, we adopt canonical spectral values of $\beta = -2.3$ and $E_{\rm peak} = 400 \rm~keV$ compatible with observation of short GRBs \citep{2011A&A...530A..21N}. This ensures a physically consistent Band-function spectral shape when only $\alpha$ is known.

Our range of $k$-corrections is consistent with values reported by \citet{rowlinson2013} and \citet{2001AJ....121.2879B}, spanning approximately 0.4 to 7. However, in some time bins of bursts, the derived k-correction is large, indicating that the spectral shape in the extrapolated frequency range is likely insufficiently constrained. We clarify that $k < 1$ in some cases (particularly for GRB 080919) arises from fitting Swift/XRT band (0.3--10 keV) spectra with steep photon indices (e.g., $\alpha \sim 2.1 - 2.4$), where extrapolation to the bolometric range does not substantially increase the total factor. To ensure compatibility, in these cases, we use a cutoff power-law model for the spectrum and set the cutoff energy at 300 keV. Moreover, $k < 1$ can be considered as an effect of using 1 keV as the lower integration limit, which, together with $\alpha \sim 2.1 - 2.4$, leads to $k < 1$. Conversely, values of $k > 1$ arise from relatively flat spectra, indicating that a significant fraction of the emission lies likely outside the observed Swift/XRT band. 

 Alternatively, one can use a set of Band-function template spectra. For each template $i$ with the Band function with parameters $(\alpha, \beta_i, E_{0,i})$, replace $\alpha_i$ by the measured $\alpha$, then compute k for each template.   The final $k$-correction csn be taken as the median of the set $\{k_i\}$, with uncertainties estimated from the scatter among them. We plan to apply this spectral modeling approach with full error propagation in future, more comprehensive studies involving larger samples.
}



Throughout this article, we adopt a $\Lambda$CDM cosmology with H$_0 = 69.6 \, \mathrm{km \, s^{-1} \, Mpc^{-1}}$, $\Omega_{\rm M} = 0.286$, and $\Omega_{\Lambda} = 0.714$ for performing $k$-corrections related to the cosmological rest frame of the sources.

\subsection{Multipolar Electromagnetic Spin‐Down}\label{subsec:model}
We model the plateau as spin‐down of a vacuum NS decomposed into vector spherical harmonics (dipole $l=1$, quadrupole $l=2$, hexapole $l=3$; \citealt{barrera1985,1998clel.book.....J, 2024ApJ...974...89W,Yorgancioglu2025Dainotti}).  The total multipolar spin-down luminosity ($L_{\rm SD}$) is the sum of all multipolar contributions
\[L_{\rm SD}(t)=\sum_{l=1}^{\infty}C_l\,B_l^2\,R^{2l+4}\,\Theta_l^2\,\Omega(t)^{2l+2},\]
where $\Omega$ is the NS angular velocity, $R$ the stellar radius, $C_{l}$ is a constant, $B_{l}$ is the magnetic field, $\Theta_{l}$ is the angular term.
Conservation of rotational energy yields \citep[see][for more information]{Yorgancioglu2025Dainotti},
\[
\dot{\Omega}=-\sum_{l=1}^{\infty}K_l\,\Omega^{2l+1},
\quad K_l=\frac{C_lB_l^2R^{2l+4}\Theta_l^2}{I}.
\]

\textbf{}

\subsection{Fitting and Model Selection}\label{subsec:stats}
For data fitting, we employed the LMFIT package \citep{2014zndo.....11813N}, a robust Python library for non-linear optimization and curve fitting\footnote{\href{https://lmfit.github.io/lmfit-py/}{https://lmfit.github.io/lmfit-py/}}. We systematically \textcolor{black}{explored} various magnetic field configurations, \textcolor{black}{starting} with a baseline dipolar \textcolor{black}{component} and incrementally adding higher-order \textcolor{black}{terms}. To evaluate these nested models, we \textcolor{black}{applied} the statistical F-test, comparing the simple dipolar model ($\rm L_{dip}$) with \textcolor{black}{more complex configurations} including quadrupole and hexapole components ($\rm L_{dip} + L_{quad} + L_{hexa}$). A p-value below 0.05 \textcolor{black}{was taken as evidence of} a statistically significant improvement in fit quality \citep{casella2002statistical}. \textcolor{black}{In cases where higher-order terms dominated, the final fit retained only those significant components.}

\textcolor{black}{Our analysis focused exclusively on the X-ray light curve segment prior to the onset of the fast drop phase, as this latter phase is interpreted as emission from the newly formed black hole rather than the spin-down of the magnetar. By restricting the fit to the plateau and gradual decay phases, where the number of counts per bin is sufficiently high, the Gaussian approximation to the Poisson distribution holds, and the assumptions of normality and approximately constant variance are reasonably satisfied. As an additional robustness check, we computed the Akaike Information Criterion \citep[AIC;][]{1974ITAC...19..716A} from the same fits, finding results consistent with the F-test preference and yielding $\Delta\mathrm{AIC} \le 2$ between the competing models \citep{doi:10.1177/0049124104268644}.
}

Table~\ref{table:List} lists best‐fit $B_l$, $\Omega_0$, and the collapse time, $t_{\rm co}$.   Pre‐collapse period is $P_{\rm pre}=P_{\rm NS}(t) = 2\pi/\Omega(t)$ for mass–spin inference (Sect.~\ref{sec:results}). The collapse time is identified when $L_{\rm SD}(t)$ steeply decays.
\bigskip

\begin{figure*}
\centering
\includegraphics[width=0.58\textwidth,clip]{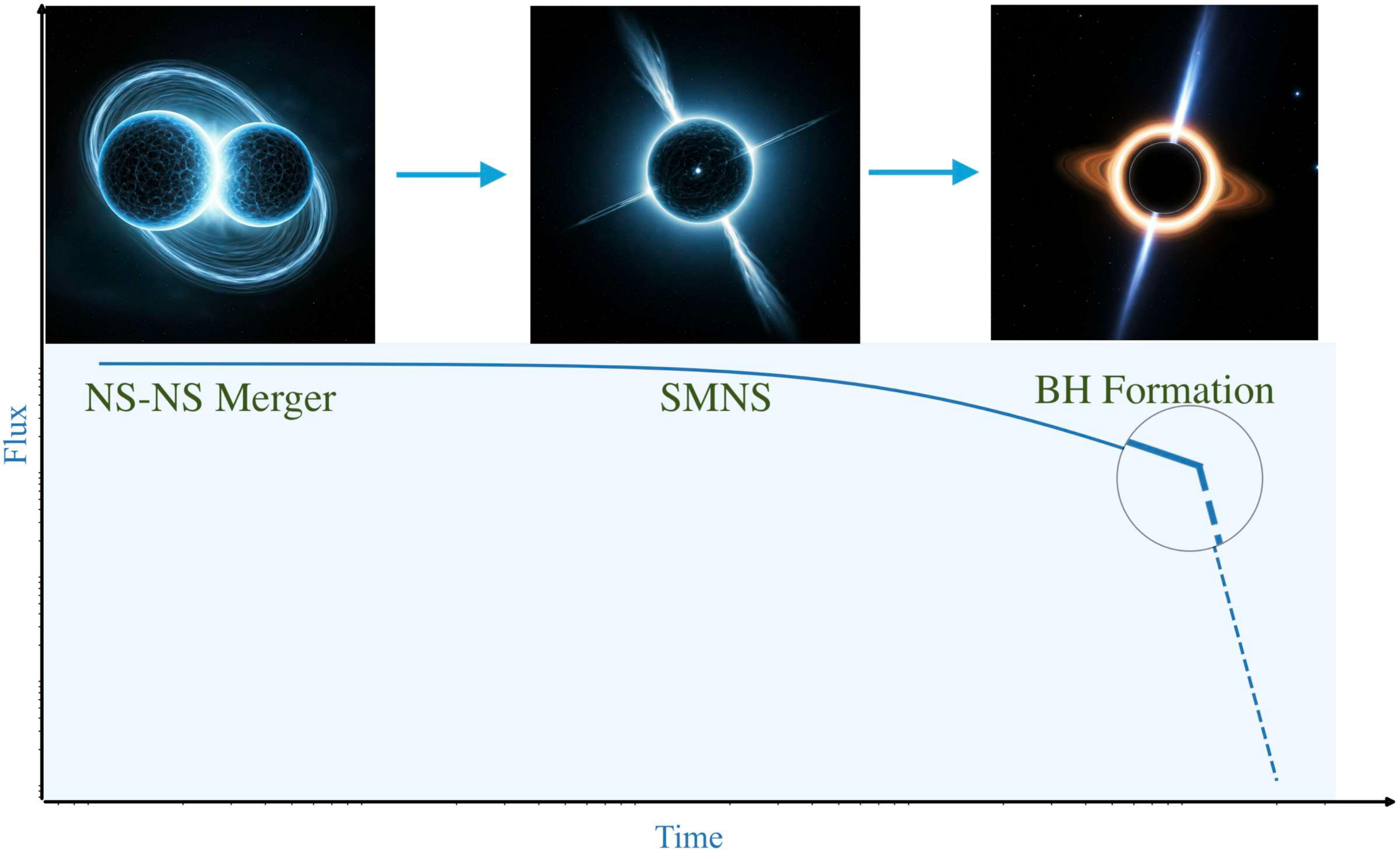}
\caption{Schematic (not from simulation): The merger of two neutron stars (NSs) is a leading model for SGRBs. If the remnant's mass is low enough, it may become a SMNS through secular spin-down. The X-ray plateau phase in SGRBs suggests magnetic spin-down from a rapidly rotating, highly magnetized SMNS. The collapse of this SMNS into a BH likely causes the steep decline in X-ray flux observed at the plateau's end.}
\label{fig:scheme}
\end{figure*}

\begin{figure*}
\centering
\includegraphics[width=0.3\textwidth,clip]{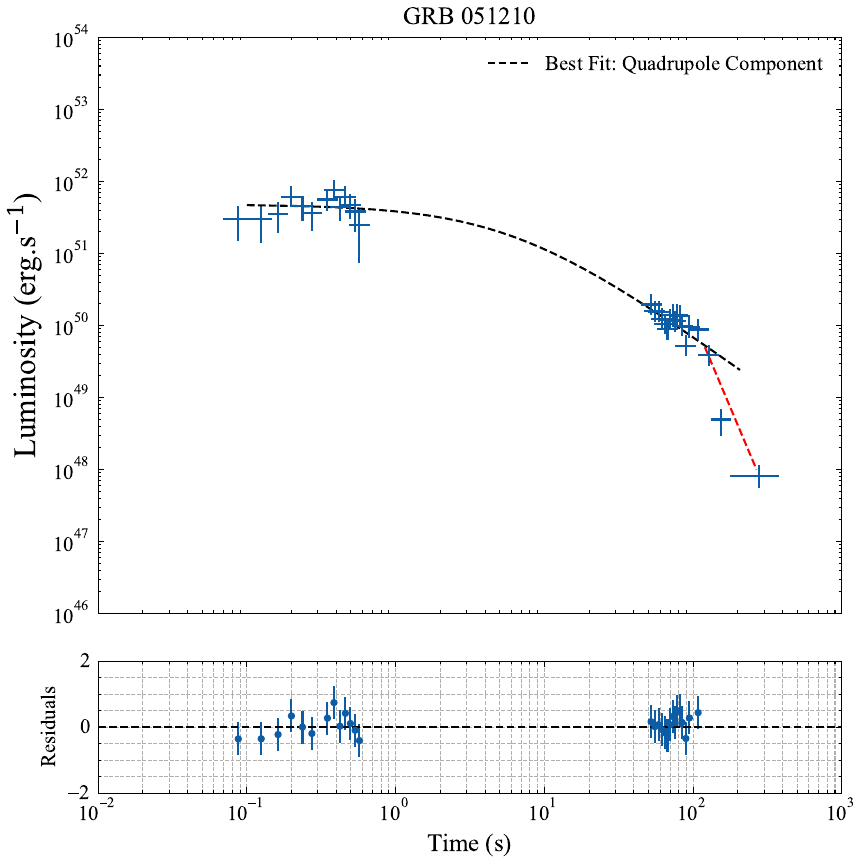}
\includegraphics[width=0.3\textwidth,clip]{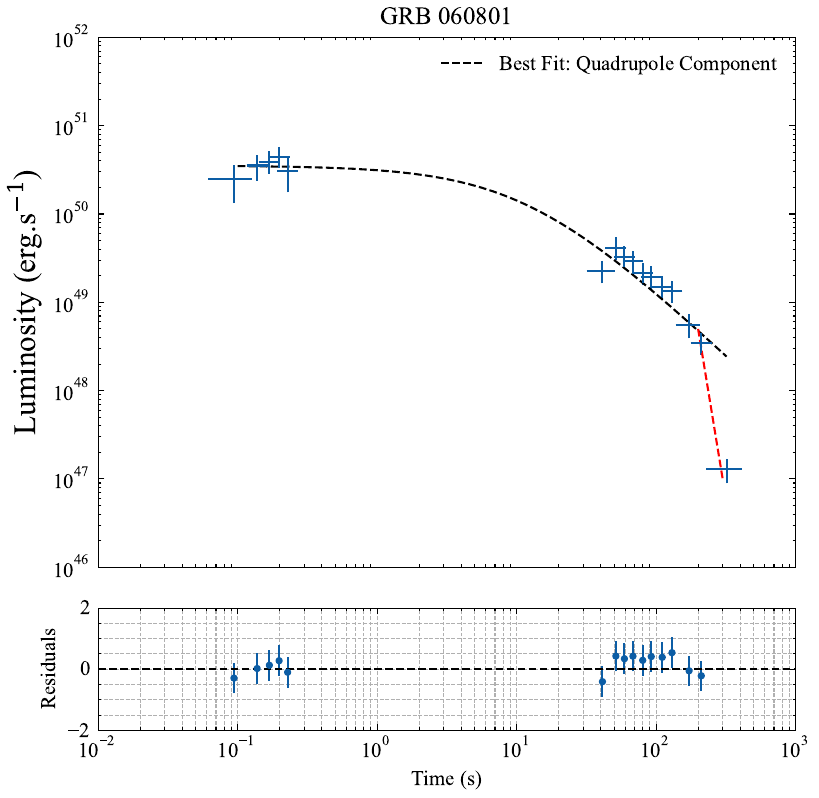}
\includegraphics[width=0.3\textwidth,clip]{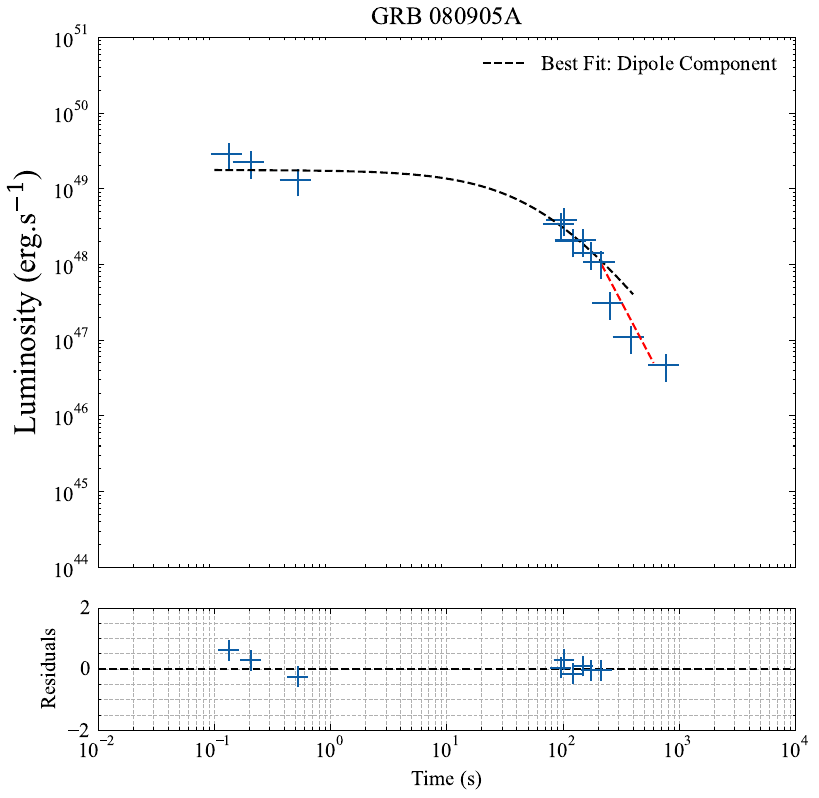}
\includegraphics[width=0.3\textwidth,clip]{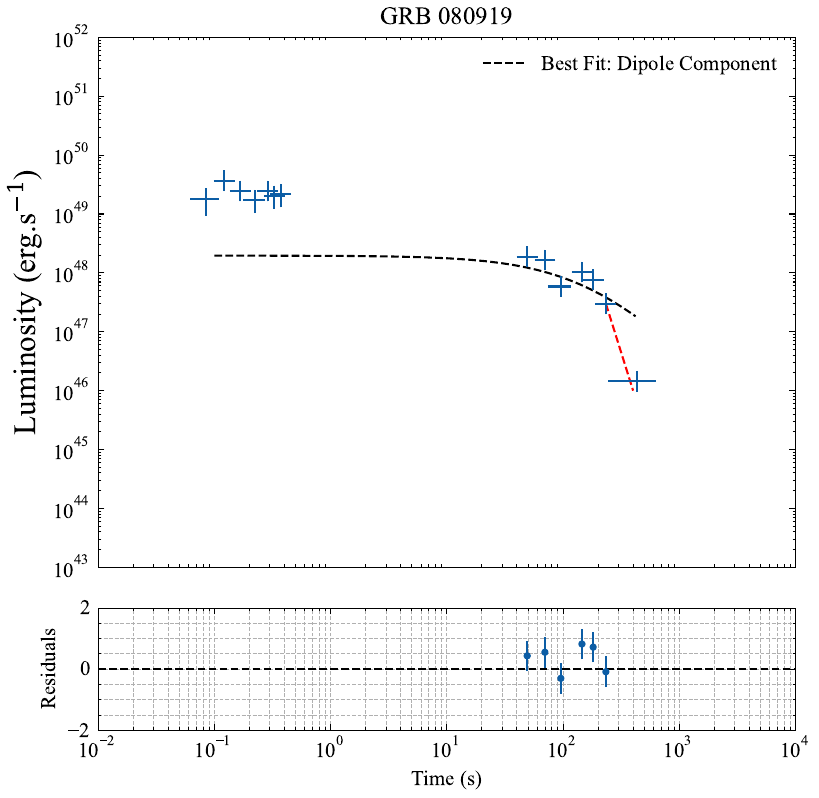}
\includegraphics[width=0.3\textwidth,clip]{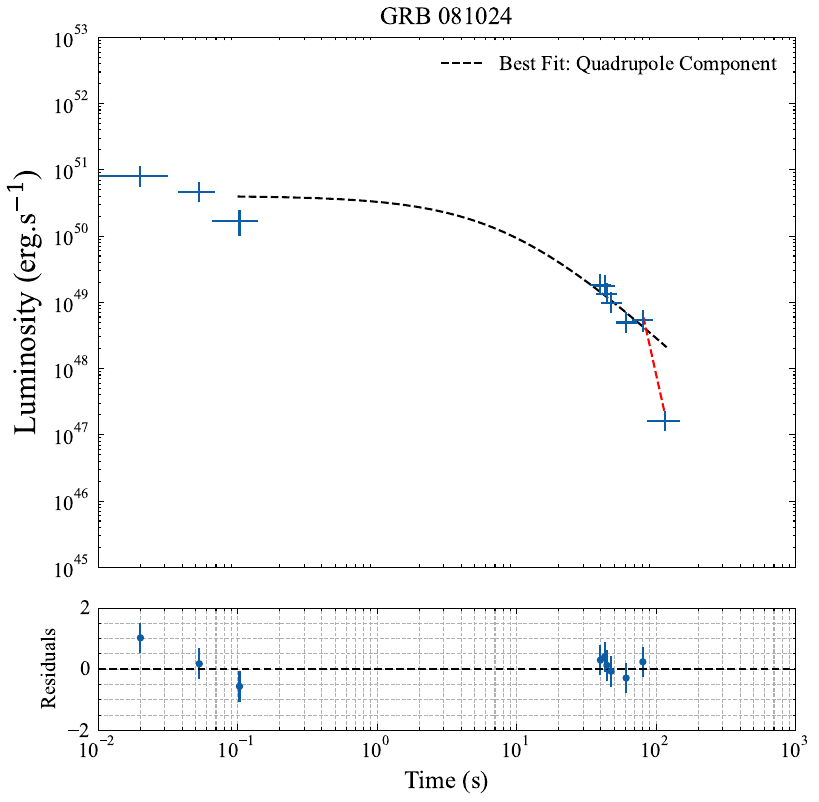}
\includegraphics[width=0.3\textwidth,clip]{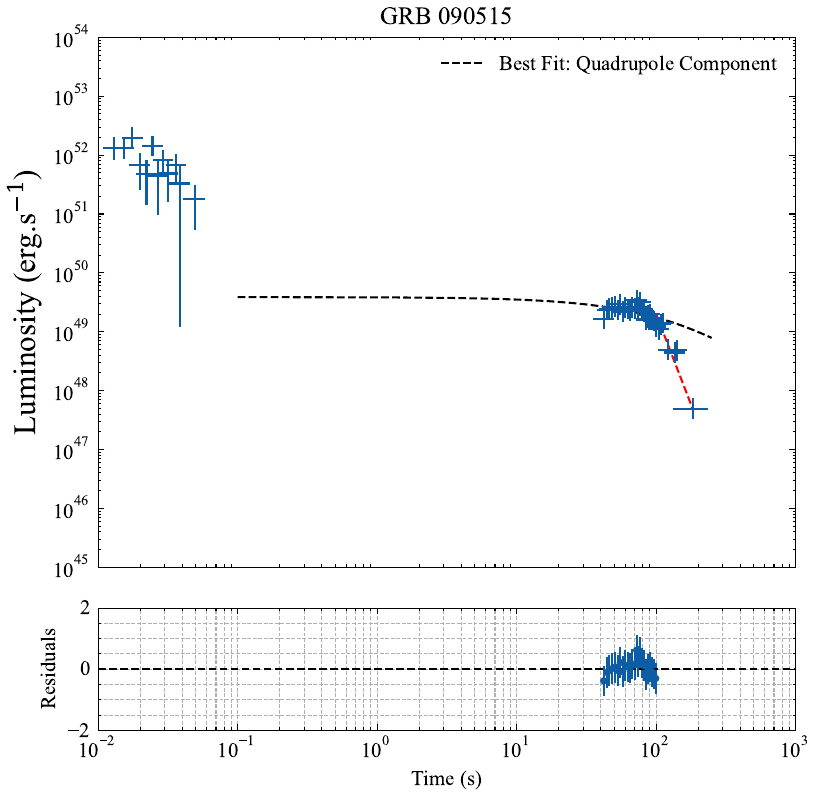}
\includegraphics[width=0.3\textwidth,clip]{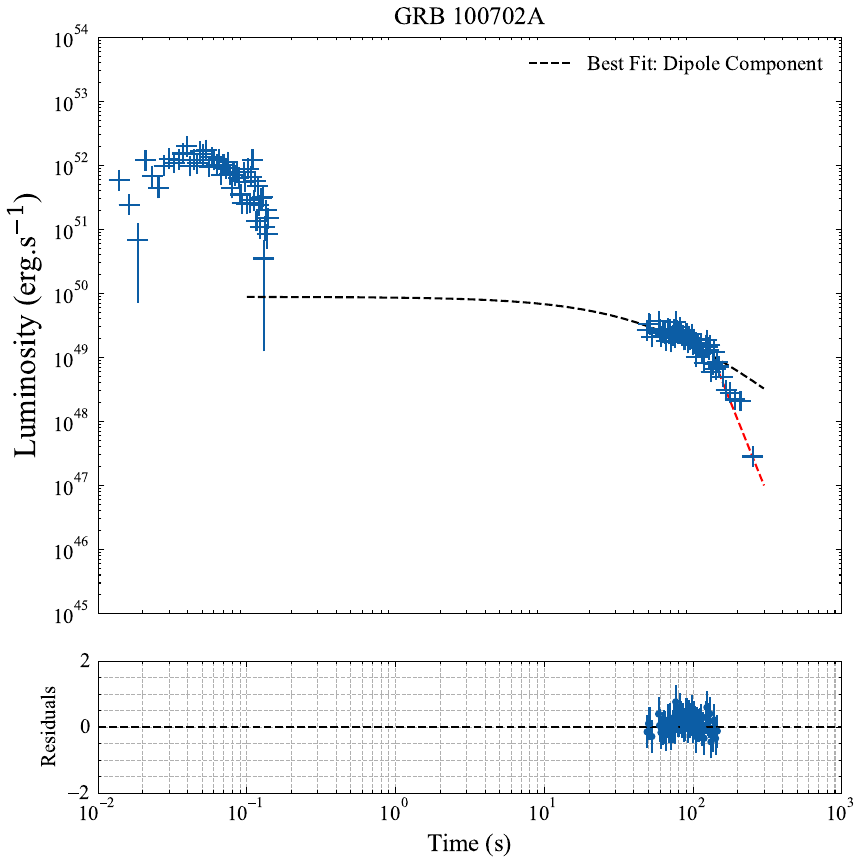}
\includegraphics[width=0.3\textwidth,clip]{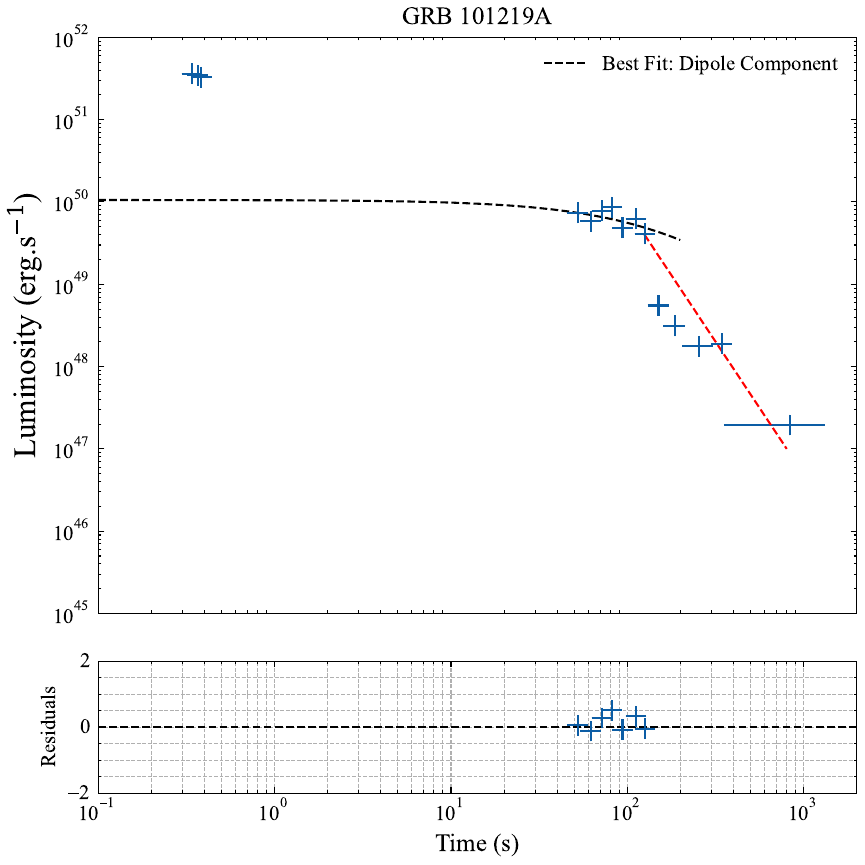}
\includegraphics[width=0.3\textwidth,clip]{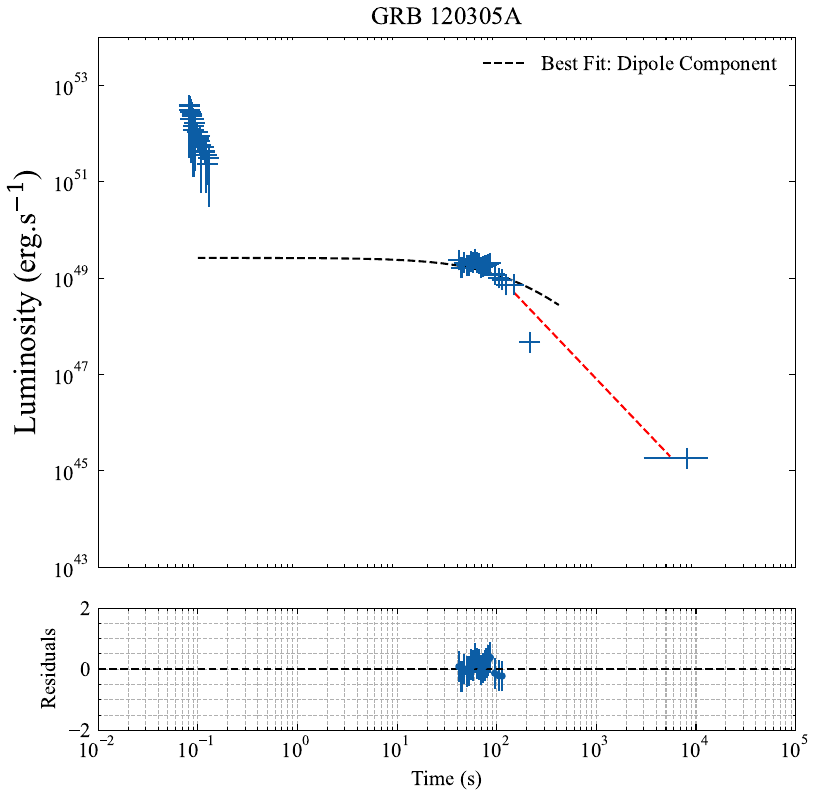}
\includegraphics[width=0.3\textwidth,clip]{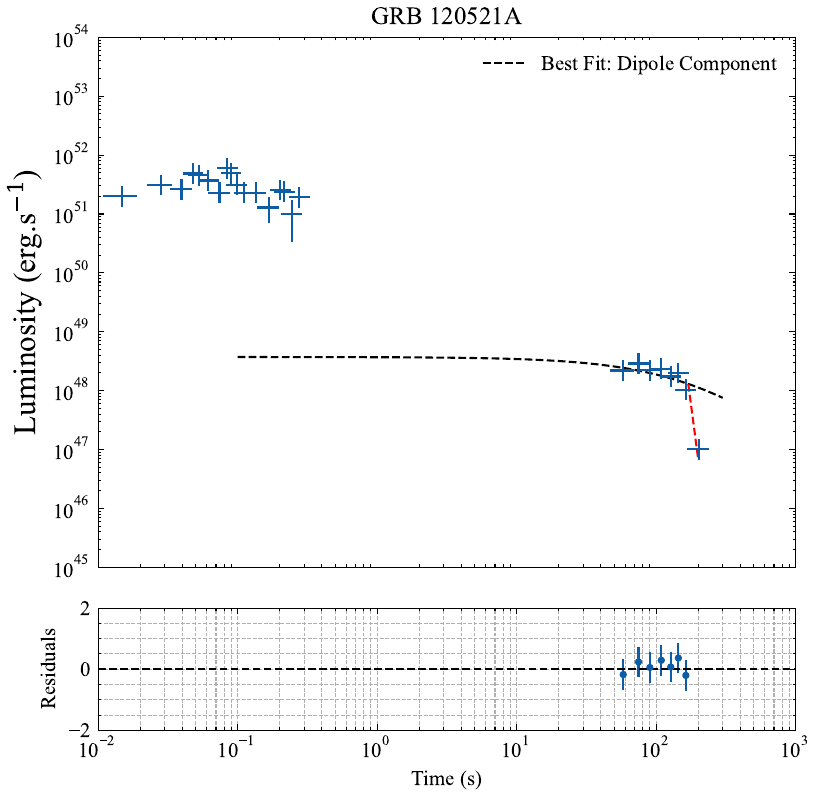}
\caption{\textbf{A:} NS spin-down energy as the primary energy source of the GRB afterglow. Initial spin period (P$_0$) and magnetic field components, are detailed in Table~\ref{table:List}. The plot is presented on a logarithmic scale, and the residuals are expressed relative to the model-predicted luminosity:
$
\text{Residual} = \frac{\text{Observed Luminosity} - \text{Model-Predicted Luminosity}}{\text{Model-Predicted Luminosity}}$.}

\label{fig:LC}
\end{figure*}

\begin{figure*}
\centering

\includegraphics[width=0.3\textwidth,clip]{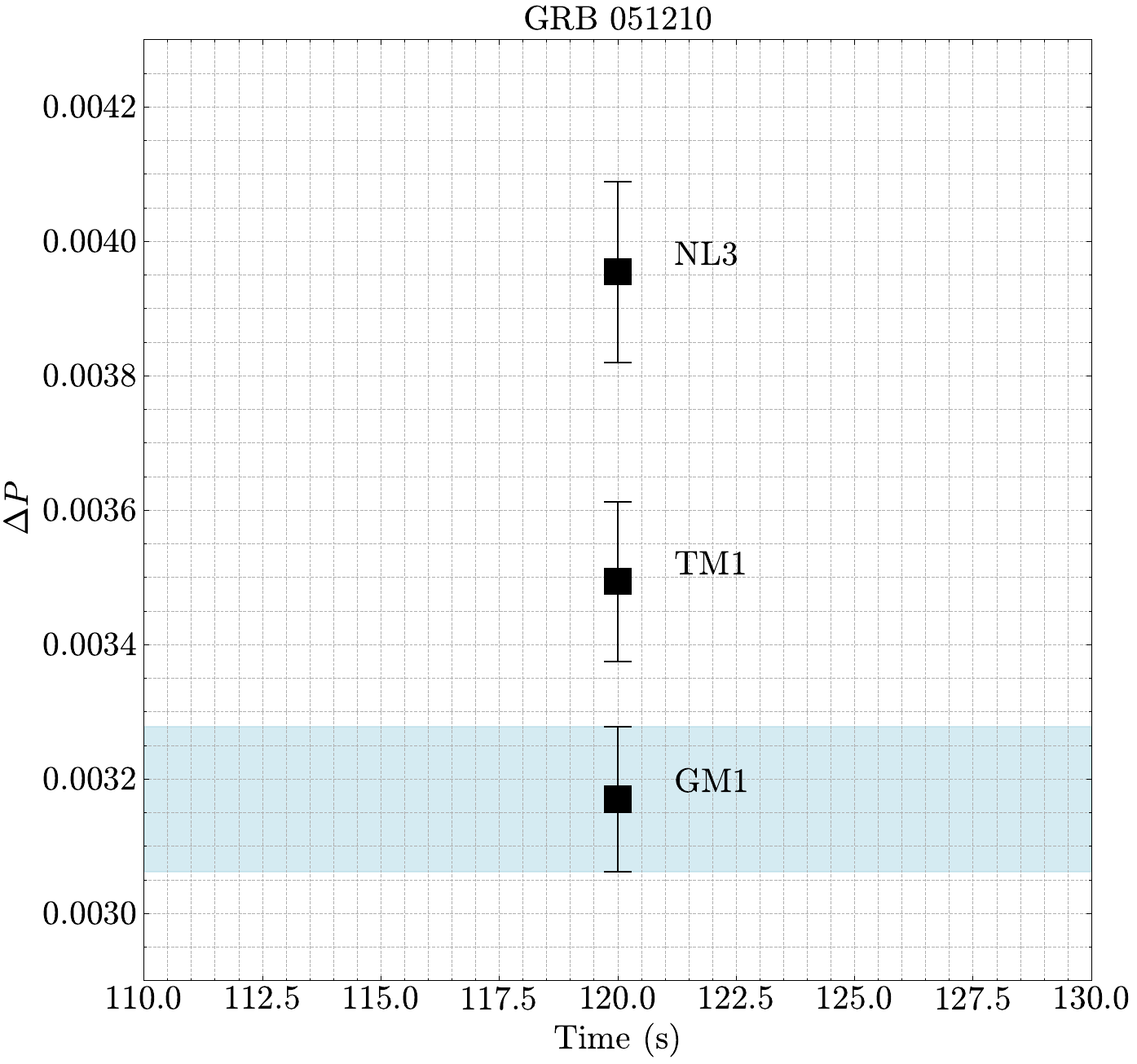}
\includegraphics[width=0.3\textwidth,clip]{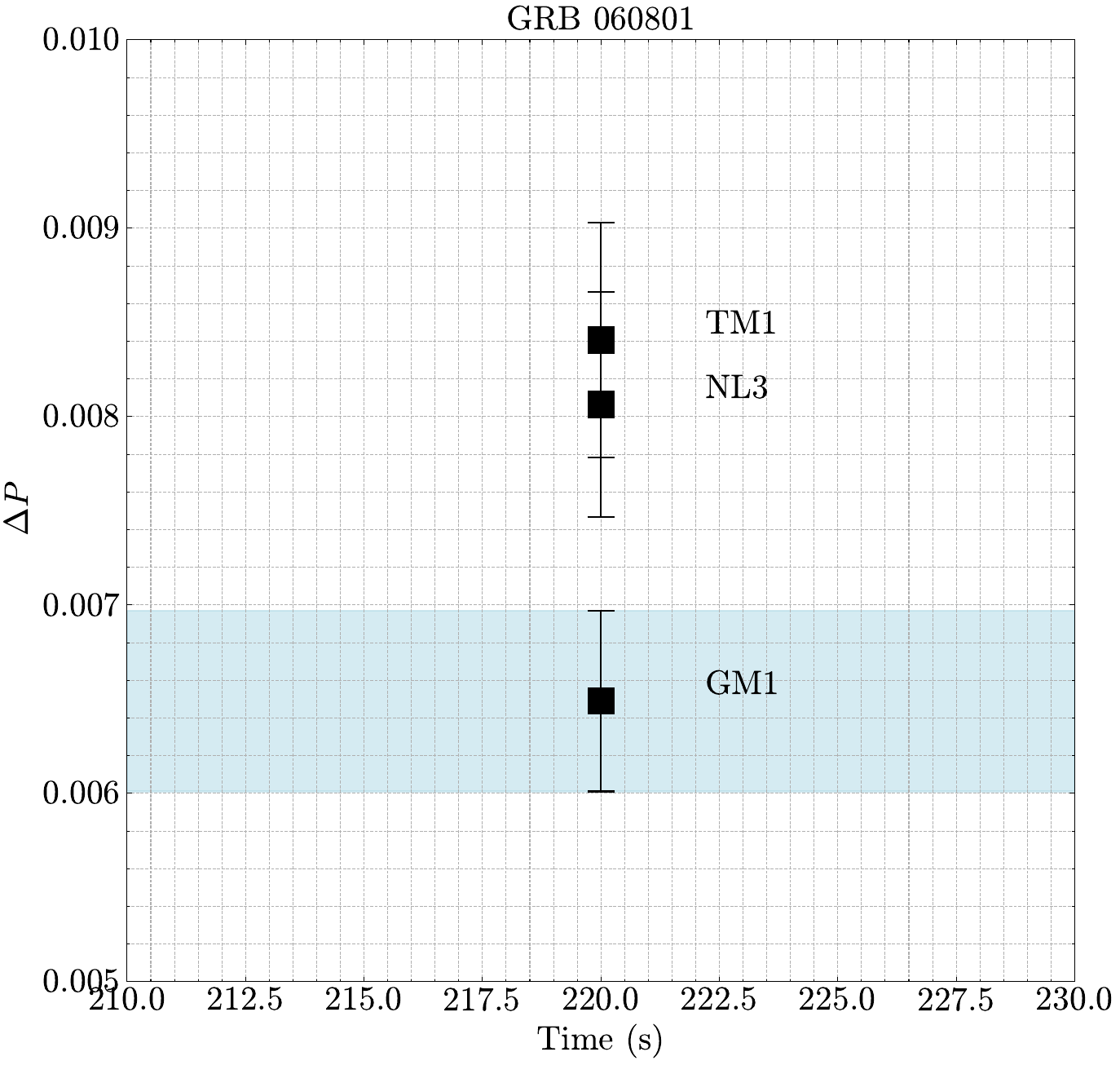}
\includegraphics[width=0.3\textwidth,clip]{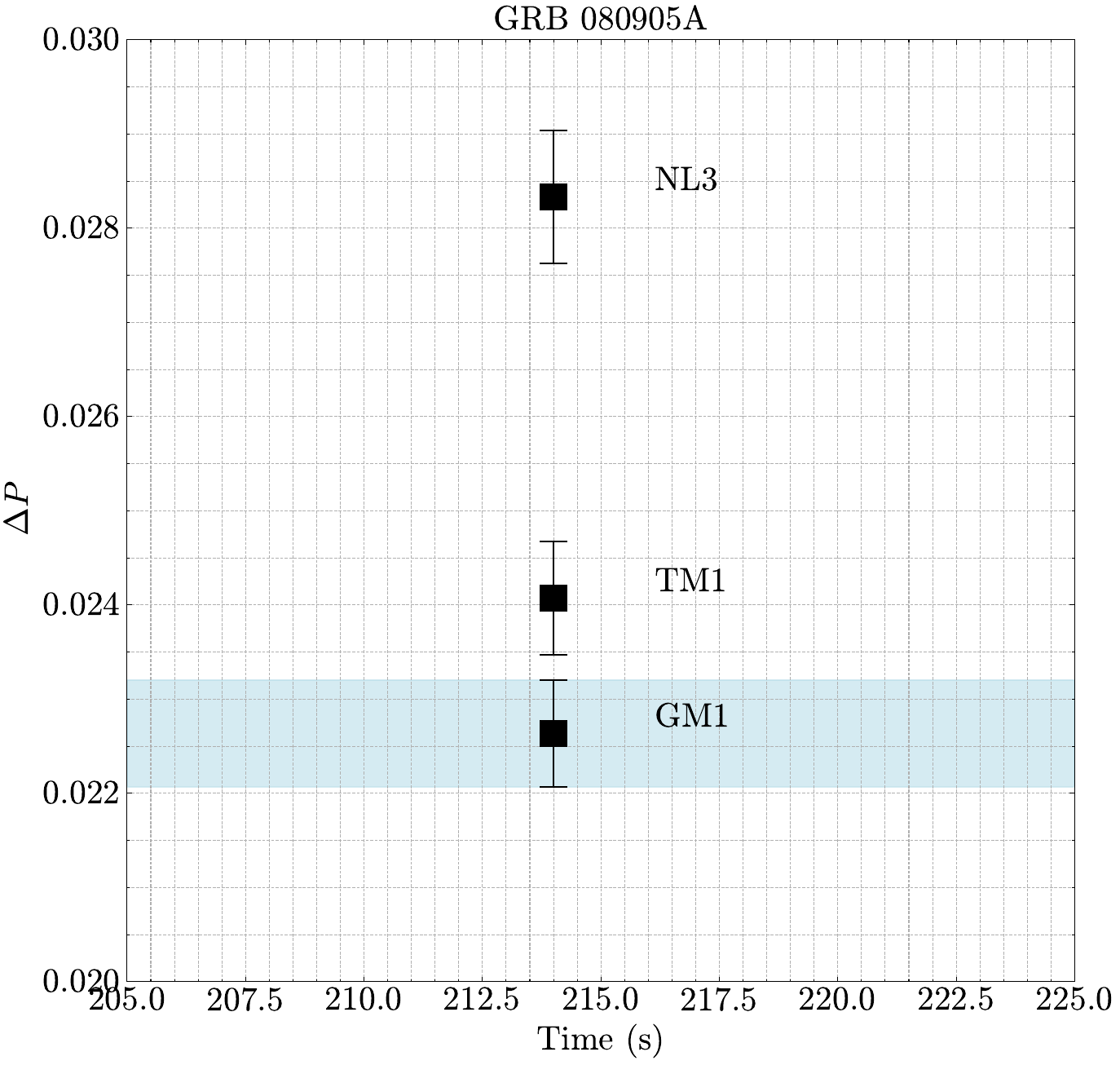}
\includegraphics[width=0.3\textwidth,clip]{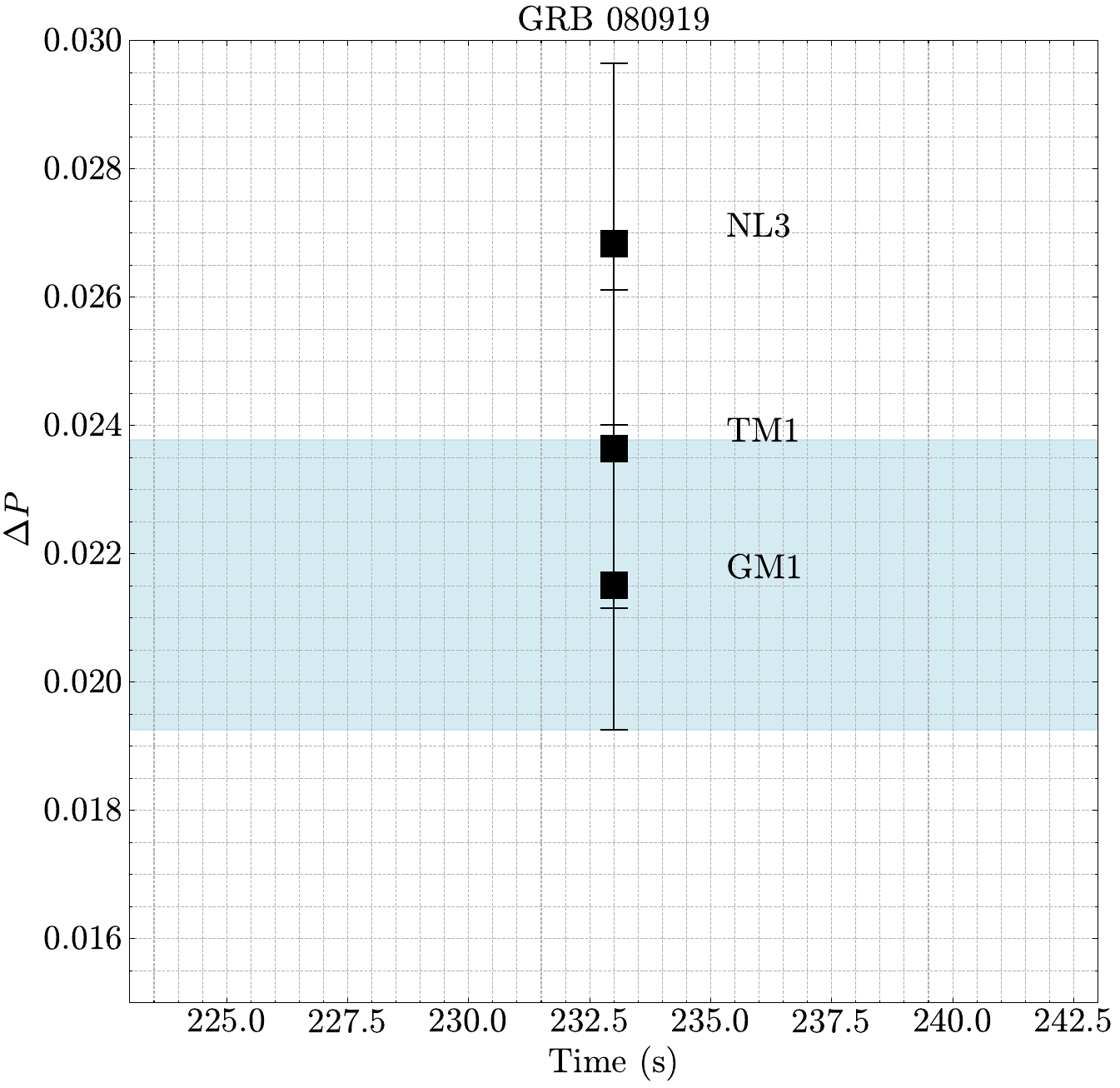}
\includegraphics[width=0.3\textwidth,clip]{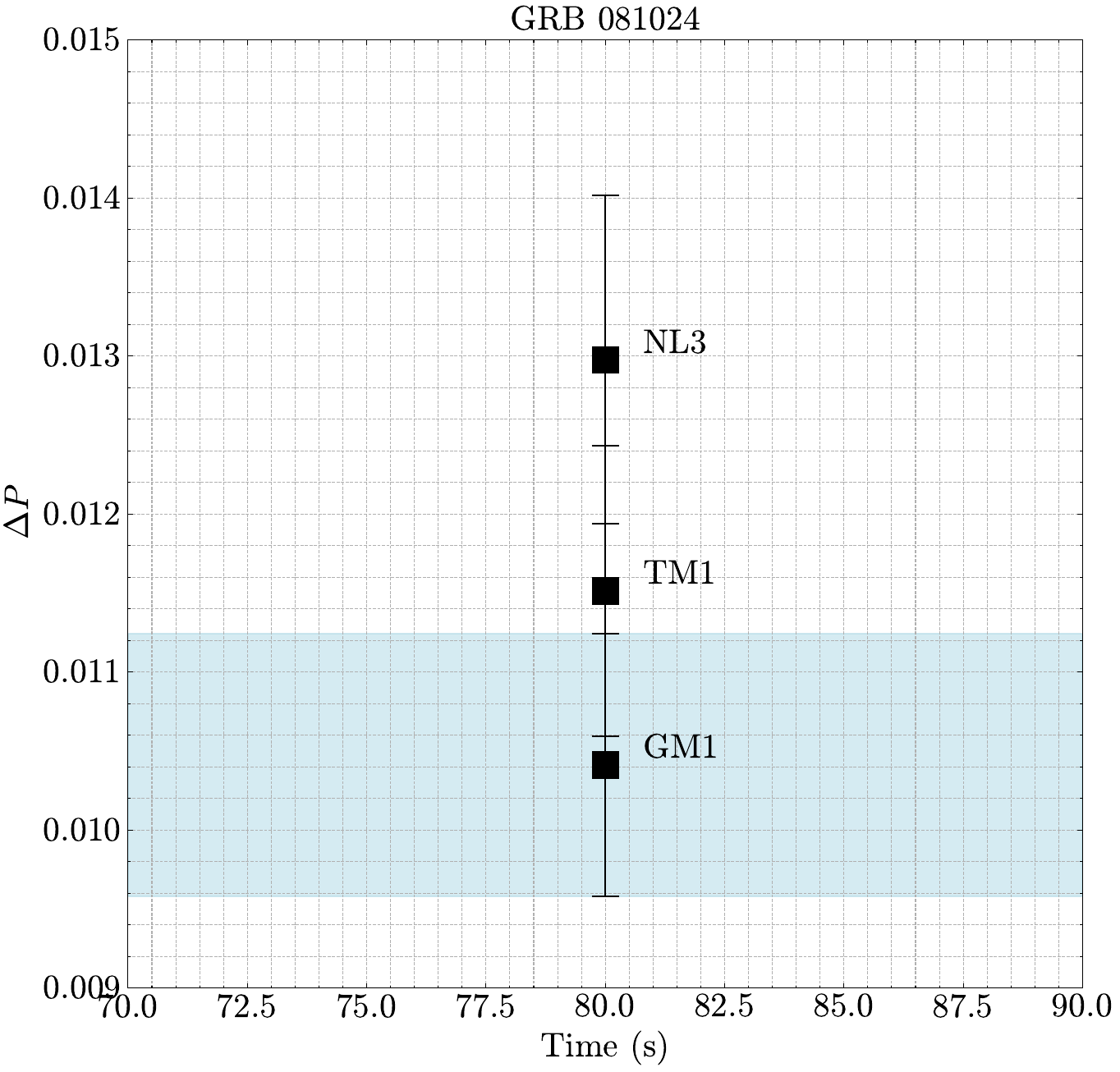}
\includegraphics[width=0.3\textwidth,clip]{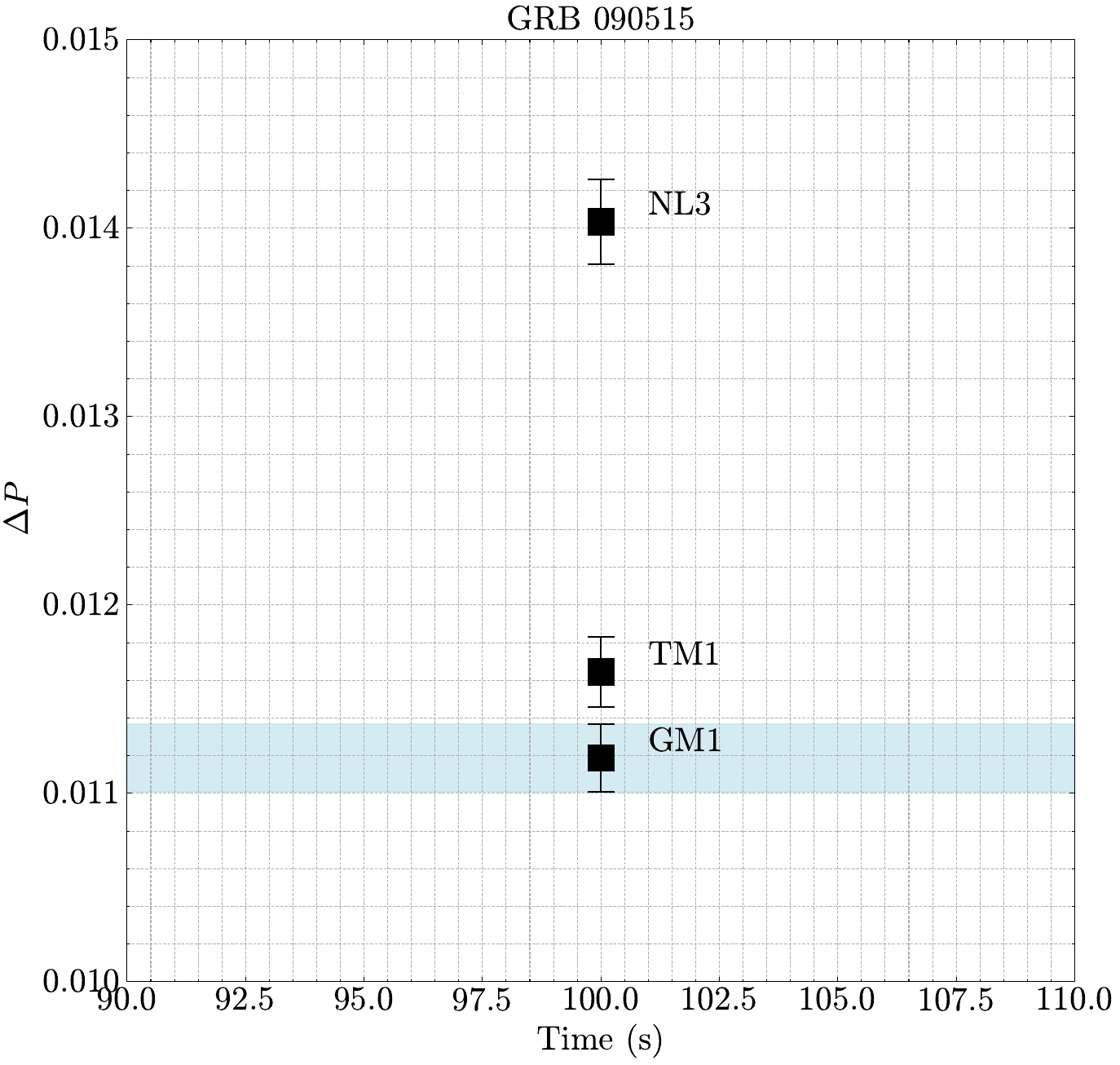}
\includegraphics[width=0.3\textwidth,clip]{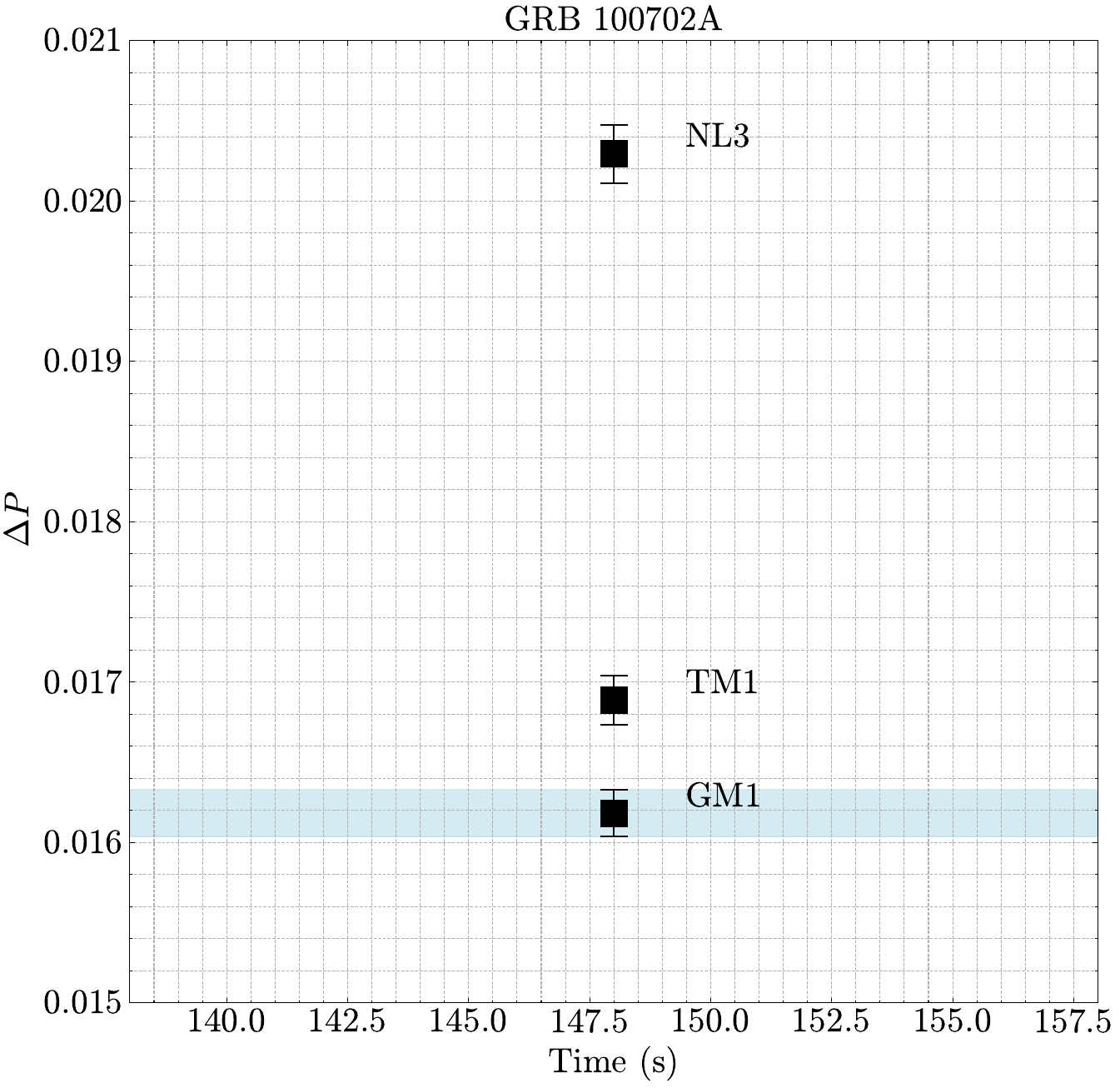}
\includegraphics[width=0.3\textwidth,clip]{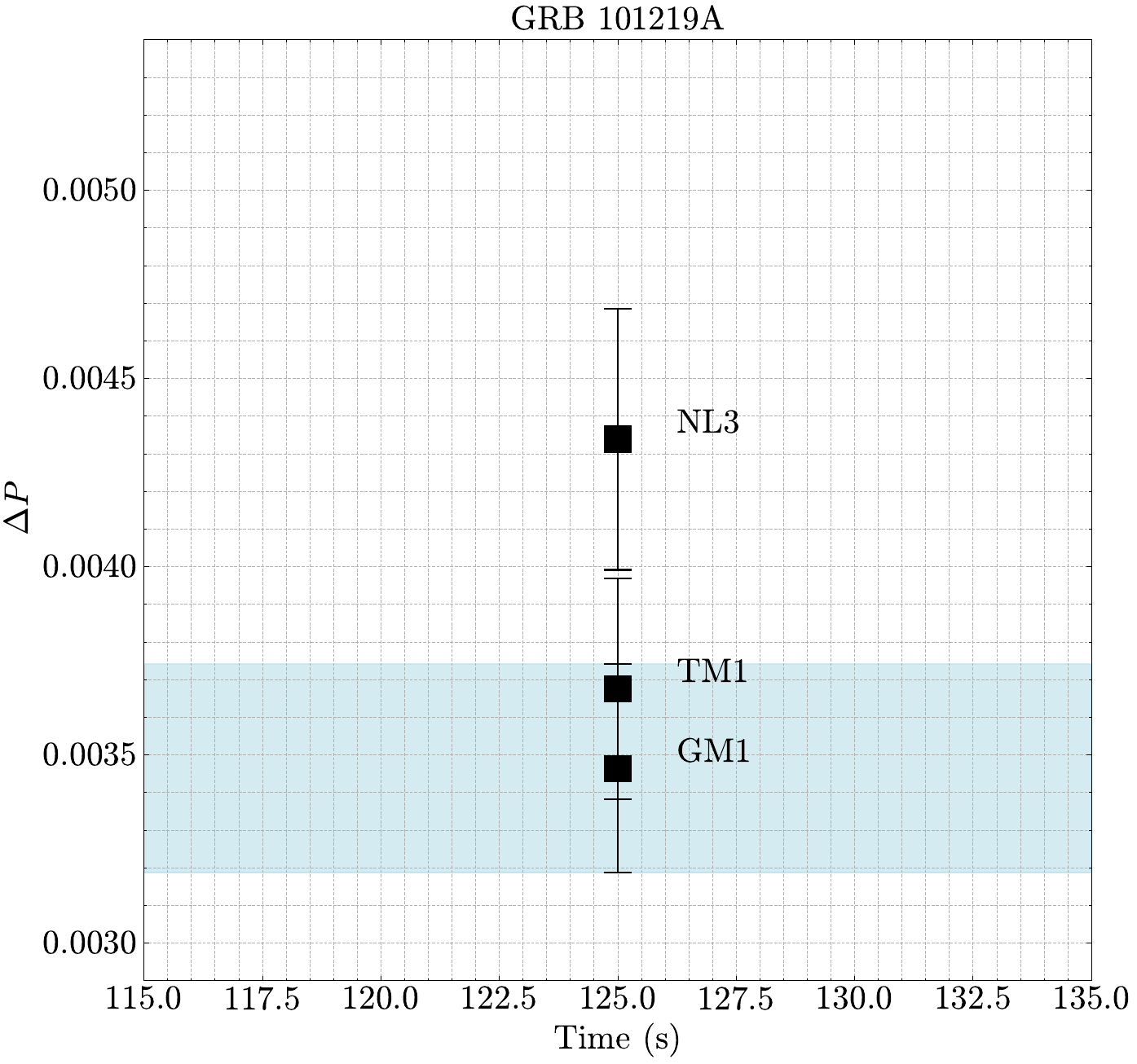}
\includegraphics[width=0.3\textwidth,clip]{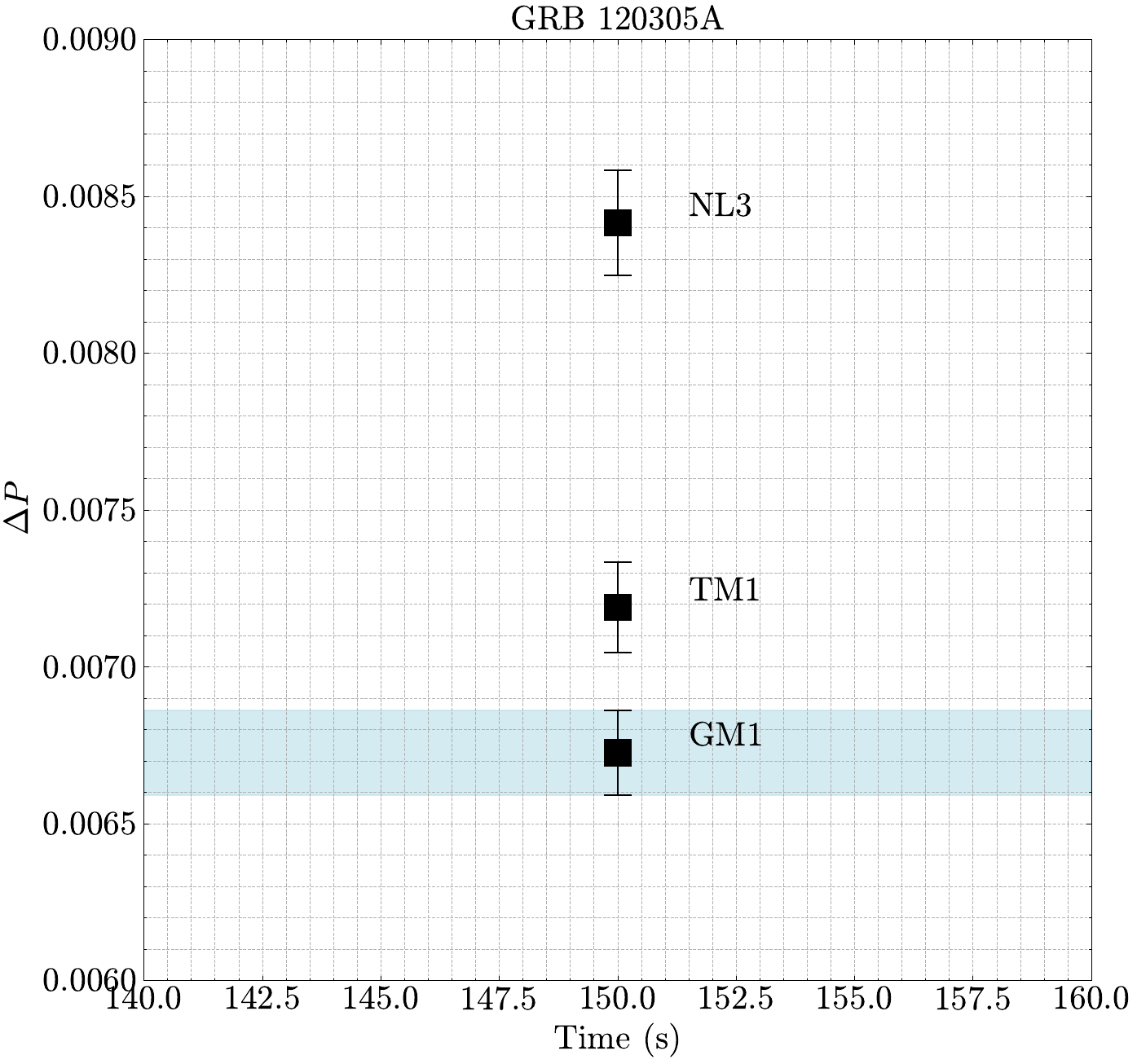}
\includegraphics[width=0.3\textwidth,clip]{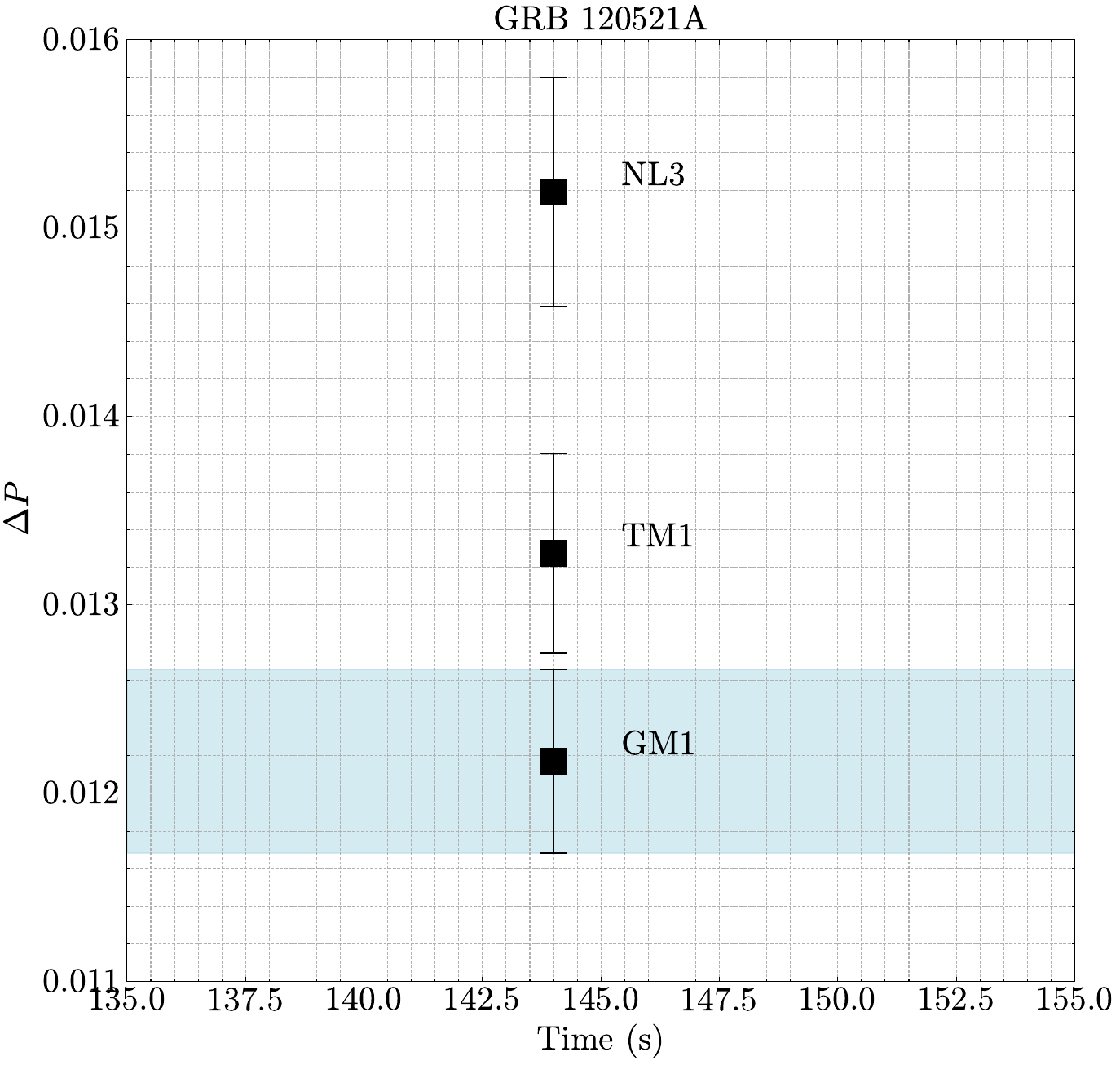}
\caption{ The GM1 EOS, representing an intermediate stiffness, provides the best fit to observational data by minimizing the absolute difference \(
      \Delta P = \bigl|\,P_{\rm NS}(t_{\rm co}) - P_{\rm BH}(t_{\rm co})\bigr|,
    \)
    where \(P_{\rm NS}(t_{\rm co})\) is the spin period of the neutron star at collapse (from multipolar spin‑down) and \(P_{\rm BH}(t_{\rm co})\) is the black‑hole spin period inferred from post‑plateau energy (see Section~\ref{sec:methods}).  GM1 yields the smallest average \(\Delta P\), with only a marginal difference from TM1 (\(M_{\rm max}=2.2\,M_\odot\)), as indicated by \(\Delta\langle\Delta P\rangle\sim10\%\).  This corresponds to \(M_{\rm crit}\approx2.3\pm0.1\,M_\odot\).  The uncertainty on \(M_{\rm crit}\) shrinks with broader temporal coverage, and when only the highest‑quality light curves are used, the best‑fit value shifts to \(M_{\rm crit}\approx2.4\,M_\odot\) with noticeably smaller error bars.}
\label{fig:080905AF}
\end{figure*}

\begin{table}[ht]
\centering
\begin{tabular}{rccccc}
\hline
EOS  & $M^{J=0}_{\rm crit}$ & k& l &R& I \\
  ~& $(M_\odot)$ & &  & (km)&  (10$^{45}$g cm$^2$) \\

\hline

TM1\ & 2.20 & 0.017& 1.61&12.5&2.81 \\

GM1 & 2.39 & 0.011&  1.69&12.1& 3.12\\

NL3 & 2.81 & 0.006&  1.68&13.8& 4.99\\
\hline
\hline
\end{tabular}

\caption{Properties of NSs for the selected EOS: TM1, GM1, and NL3. The table lists the critical mass for the non-rotating case, \(M^{J=0}_{\rm max}\), along with the parameters \(k\) and \(l\) introduced in Eq.~\ref{eq:Mcrit}, as well as the radius and moment of inertia for each EOS.\label{tab:propNS}}

\end{table}

\begin{table*}[ht]
\centering
\footnotesize  
\begin{tabular}{lcccccccc}
\hline

SGRB & z	& E$_{\rm iso}$ & P$_{0}$ & B$_{\rm dip}$ & B$_{\rm quad}$ & Collapse time & $E_{\rm PD}$  & $\Delta P$ \\       
      &  & (erg)        & (ms)      & ($10^{15}$ G)         &     ($10^{15}$ G)                             & (s)           & (erg )   & (ms)     \\
\hline
051210 & (0.72) & 5.98$^{+13.5}_{-4.05}\times$10$^{51}$ & 1.01$\pm 0.29$   & --    & 13.7$\pm 0.9$                & 120$\pm$13 & (1.68$\pm$ 0.44)$\times$10$^{51}$ & $3.2\pm 0.1$ \\
060801 & 1.13 & 1.17$^{+1.79}_{-0.71}\times$10$^{53}$ & 2.7$\pm 0.2$   & -   & 69.8 $\pm 0.9$               & 220$\pm$24 & (2.68$\pm0.74$)$\times$10$^{50}$ & $6.4\pm 0.5$ \\
080905A &0.122 & 6.16$^{+12.3}_{-4.03}\times$10$^{50}$ & 6.9$\pm 0.2$  & 18.7$\pm 0.2$ & -  & 214$\pm$ 12 & (6.1 $\pm 0.41$)$\times$10$^{49}$& $22.6\pm 0.6$ \\
080919 & (0.72) & 5.18$^{+9.34}_{-3.26}\times$10$^{51}$ & 15.1$\pm 2.3$  & 21.6$\pm 1.1$ & -                     & 233$\pm$35 & (3.91$\pm1.01$)$\times$10$^{49}$ & $21 \pm 3$ \\
081024 & (0.72) & 5.65$^{+7.53}_{-3.16}\times$10$^{51}$ & 5.2$\pm 0.4$   & 32.5$\pm 0.4$   & -                   & 80$\pm$10 & (1.5$\pm$0.24)$\times$10$^{50}$  & 10$\pm$0.6 \\
090515 & (0.72) & 3.44$^{+3.55}_{-1.55}\times$10$^{50}$ & 2.8$\pm 0.1$   & 3.6$\pm0.3$   & -   &                  100$\pm$3 & (3.8$\pm$ 0.81)$\times$10$^{50}$ & 11$\pm$0.1  \\
100702A & (0.72)& 2.28$^{+1.46}_{-0.80}\times$10$^{51}$ & 2.9$\pm 0.1$   & 6.8$\pm 0.7$   & -  &                   148$\pm2$ & (1.72$\pm$0.37)$\times$10$^{50}$ & 16$\pm$0.1  \\
101219A& 0.718 & 1.69$^{+0.79}_{-0.54}\times$10$^{53}$ & 1.6$\pm 0.1$   & 1.9$\pm 0.2$    & -  &125$\pm$13 & (2.59$\pm$0.28)$\times$10$^{51}$ & 3.4$\pm$0.3 \\
120305A &(0.72) & 2.02$^{+0.10}_{-0.10}\times$10$^{52}$ & 3.7$\pm$0.1   & 6.4$\pm 0.1$    & -  &150$\pm$27 & (6.17$\pm$1.2)$\times$10$^{50}$& 6.7$\pm$0.1 \\
120521A & (0.72)& 8.42$^{+12.19}_{-4.95}\times$10$^{51}$ & 9.2$\pm 0.3$   & 11.4$\pm 0.5$    & -  & 144$\pm40$ & (1.36$\pm$0.39)$\times$10$^{50}$ &12.1$\pm$0.4\\
\hline
\hline
\end{tabular}
\caption{The table presents the sample of SGRB magnetars along with their best-fit parameters for GM1 EOS. The redshift, z, and the isotropic equivalent energy, E$_{\rm iso}$, are taken from \cite{rowlinson2013}. When the redshift is unknown, an average value of (0.72) is used. Parameters for the initial spin period P$_{0}$, dipole magnetic field B$_{\rm dip}$, and quadrupole magnetic field B$_{\rm quad}$ are obtained by fitting the model, assuming isotropic emission. The collapse time is considered to coincide with the onset of the steep decay phase observed in SGRB afterglows. The post-decay energy, $E_{\rm PD}$, is determined by multiplying the luminosity by the time interval for each data point following the decay in the luminosity light curve, and then summing these products over the entire observation period. The size of the uncertainties is inversely correlated with the quantity of data available for fitting: more extensive coverage yields smaller error bars. Consequently, higher‑quality data produce more precise and reliable results, as is evident in our dataset.
 }
\label{table:List}
\end{table*}

\section{Results}\label{sec:results}

\cite{rowlinson2013} analyzed all Swift SGRBs observed up to May 2012, finding $\sim$50\% (10 SGRBs) exhibit clear magnetar plateau phases. While the remaining cases are consistent with magnetar formation, data limitations preclude definitive plateau confirmation. We study this 10 SGRBs with X-ray plateaus and steep decays, reproducing their bolometric (1--10000 keV) light curves (Section~\ref{sec:methods}). Critically, rather than assuming pure dipole spin-down, we incorporate recent evidence for multipolar magnetic fields \citep{2024ApJ...974...89W} (Section~\ref{sec:methods}). For some SMNS candidates, higher-order components (e.g., quadrupole) yield statistically robust fits, with collapse times aligning with multipole-dependent spin-down timescales \citep{beniamini2021}; see Fig.~\ref{fig:LC} and Table~\ref{table:List}.

As discussed in section~ \ref{sec:1}, our central engine interpretation assumes the plateau originates from magnetar multipolar electromagnetic radiation.  Traditional external shock models \citep{meszarosrees93,meszarosrees97,sari98} struggle to explain prolonged plateaus observed in $\sim$37\% of SGRBs \citep{2011A&A...526A.121D,2024A&A...692A..73G}. Although scenarios considering continuous energy injection into external shocks could produce plateaus without invoking multipolar fields \citep{2015ApJ...805...13L,2019ApJS..245....1T}, both observational and theoretical evidence suggest sustained energy injection, likely from magnetar spin-down \citep{2007ApJ...665..599T, 2011A&A...526A.121D, 2024ApJ...974...89W,2024A&A...692A..73G}.  In particular, multipolar spin-down can explain: (i) the distribution of X-ray afterglow decay indices in 238 GRBs \citep{2024ApJ...974...89W}, (ii) the Dainotti correlation, independent of multipole order \citep{Yorgancioglu2025Dainotti}, and (iii) ``internal plateaus'' and ``sudden abrupts'' inconsistent with standard afterglow models \citep{2007ApJ...665..599T,rowlinson2013}.


 In the context of SGRBs, as assumed, the collapse of ``protomagnetar'' into a KBH causes a steep decay in X-ray flux; see Figure~\ref{fig:scheme} and Figure~\ref{fig:LC}. We require that the post-collapse X‑ray energy budget can, in principle, be supplied by the rotational energy of the newly formed KBH. We do not specify a particular extraction mechanism. By the time of the plateau, any fallback or accretion disk is likely depleted, so we assume negligible accretion. Pure BH spin extraction can operate under such conditions—via Blandford–Znajek–type processes with residual magnetic flux threading the BH \citep{1977MNRAS.179..433B,2015ApJ...804L..16K,2015MNRAS.453L...1N,2018MNRAS.475..266Z,2004MNRAS.350..427K, 2009MNRAS.397.1153K}. We remain agnostic regarding the specific configuration or efficiency: our approach requires only that extraction is ``\textit{energetically}'' feasible without invoking active accretion.

 Therefore, the mass of the KBH must satisfy the condition:  \(M \geq M_{\rm crit}(\alpha),\) at the moment of collapse ($t_{\rm co}$), where $M_{\rm crit}(\alpha)$ is the EOS-dependent critical mass for a NS with dimensionless spin $\alpha$. By applying a fully relativistic treatment, \cite{2015PhRvD..92b3007C} showed that for various EOS models, such as NL3, GM1, and TM1, the critical mass of a rigidly rotating neutron star can be approximated by:
\begin{equation}\label{eq:Mcrit}
M_{\rm crit}(j) = M_{\rm crit}^{J=0} \left( 1 + k j^l \right),
\end{equation}
where \(k\), \(p\), and \(M_{\rm crit}^{J=0} = M_{\rm TOV}\) are constants that depend on the EOS, and \(j = c J /(G  M_\odot^2)\) represents the dimensionless spin parameter. By employing the relation \(j = \alpha \left( \frac{M}{M_\odot} \right)^2\), Eq.~(\ref{eq:Mcrit}) transforms into an implicit non-linear algebraic equation for the critical mass of the neutron star as a function of \(\alpha\). The values for \(M^{J=0}_{\rm max}\) can be found in Table \ref{tab:propNS}, while \(k = [0.017, 0.011, 0.0060]\) and \(l = [1.61, 1.69, 1.68]\) correspond to the EOS TM1, GM1, and NL3, respectively. This analytic solution has been obtained for values of mass along the secular axisymmetric instability line with respect to fits for each EOS, with maximum relative errors of 0.33\%, 0.44\%, and 0.45\% for the EOS TM1, GM1, and NL3, respectively.

We note that the merger remnant may initially be differentially rotating or thermally supported, but over the $\sim10$–$10^3\,$s plateau timescale such effects decay, making it appropriate to adopt the uniform‐rotation $M_{\rm crit}$ of \cite{2015PhRvD..92b3007C}.

The energy conditions for a KBH are derived from its mass-energy relation, given by:
\begin{equation}
\label{eq:mass_energy}
M^2 = \frac{c^2 J^2}{4 G^2 M^2_{\rm irr}} + M_{\rm irr}^2,
\end{equation}
where $M$ is the total mass of the black hole, $J$ is its angular momentum, $M_{\rm irr}$ is the irreducible mass, $c$ is the speed of light, and $G$ is the gravitational constant \citep{1970PhRvL..25.1596C,1971PhRvD...4.3552C,1971PhRvL..26.1344H}.

The extractable energy from a Kerr BH, $E_{\rm ext}$, is the difference between the total mass $M$ and the irreducible mass $M_{\rm irr}$:
\begin{equation}
\label{eq:Eextr}
E_{\rm ext} = (M - M_{\rm irr}) c^2= f(\alpha) M c^2,
\end{equation}
where, $f(\alpha) = \left( 1 - \sqrt{\frac{1 + \sqrt{1 - \alpha^2}}{2}} \right)$, and $\alpha \equiv cJ/(GM^2)$ is the dimensionless BH spin parameter.

Assuming that the post-decay plateau (PD) luminosity originates from the Kerr BH, we have \(
E_{\rm ext} = E_{\rm PD} = \sum_{t_i = t_{\rm co}}^{} L_{\rm PD}(t_i)  \Delta t_i, \)
where $i$ indexes each data point and $\Delta t_i$ represents the time interval between successive observations. We solve for $M$ and $\alpha$ at collapse time $t_{\rm co}$. Our aim is to show that the extractable energy of a Kerr BH (Eq.~\ref{eq:Eextr}) can power the observed post‐plateau emission, thereby constraining the BH mass $M$ and spin $\alpha$.  To do so we compute $E_{\rm PD}$ for each GRB in Table~\ref{table:List}, and then solve simultaneously \(
E_{\rm PD}=E_{\rm extr}(M,\alpha)\,, \) and \(M\ge M_{\rm crit}(\alpha)\) (from Eq.~\ref{eq:Mcrit}) at $t=t_{\rm co}$ (the onset of rapid luminosity decay), yielding a lower bound on $M$ and an upper bound on $\alpha\,$.

 Given the value of $\alpha$ and the mass $M$ at the moment of collapse, we can use the relation \(P = \frac{2 \pi I}{J} = \frac{ c}{G }\frac{2 \pi I }{ \alpha M^2 }\), to obtain the spin period $P$, which for distinguishing it from the spin period obtained from the magnetic spin dpwn formulae of the NS, $P_{\rm NS}(t_{\rm co})$, we call it $P_{\rm BH}(t_{\rm co})$.

To determine which EOS best fits the observational data, we compare the spin parameter derived from the NS's spin evolution at \( t = t_{\rm co} \), obtained using the multipolar luminosity evolution formulated in Section~\ref{sec:methods}, with the values of $P_{\rm BH}(t_{\rm co})$ for each GRB derived from the method described above. The EOS that pminimizing $\Delta P = |P_{\rm NS}(t_{\rm co}) - P_{\rm BH}(t_{\rm co})|$ is considered as the best fit. As shown in Figure~\ref{fig:LC}, Figure~\ref{fig:080905AF} and Table~\ref{table:List}, the GM1 EOS, representing an intermediate stiffness, provides the best fit for 10 GRBs.


For all 10 GRBs (Table \ref{table:List}, Fig.~\ref{fig:LC}, Fig.~\ref{fig:080905AF}), the intermediate-stiffness GM1 EOS ($M_{\rm max} = 2.39 M_\odot$) yields the smallest $\Delta P$ (Table \ref{table:List}). Regarding the errors, 1) $\Delta P$ uncertainties propagate from $L_{\rm PD}$ measurements and multipolar spin-down best fit ($\sim2\%$--$20\%$), 2) EOS‐fitting uncertainties in the BH and NS mass formulae ($\lesssim$0.45\%) are neglected, since the condition $M\ge M_{\rm crit}$ already provides a lower bound on $M$ and an upper bound on $\alpha\,$. Moreover, as indicated in Table~\ref{table:List}, when the redshift of a GRB is unknown, an mean value of 0.72 is used \citep[see][]{rowlinson2013,2022MNRAS.515.4890O}. Upon careful examination, we found that the results are not sensitive to the choice of redshift. \textcolor{black}{To validate this, we performed a dedicated test on GRB~051210 an indicative example, whose redshift was assumed to be $z=0.72$, by recalculating all relevant quantities for a range of redshifts. We have tested the sensitivity of our results by varying the redshift for GRB 051210 over the range $0.1 < z < 2$, which is consistent with the observed redshift distribution of short GRBs \citep{2022MNRAS.515.4890O}. We found that the results remain unchanged for $0.4 < z < 1.6$, while for $z < 0.4$ the TM1 EOS performs equally well or slightly better than GM1. For $z > 1.6$, the initial spin is too low leading to a dimensionless angular parameter, larger than the well-established upper limit for the NS stability, $[c J/(G M^2)]_{\rm max}\approx 0.7$ \citep{2015PhRvD..92b3007C}. This robustness likely arises because all redshift-dependent parameters—such as luminosity (used to infer the spin and magnetic field prior to collapse), collapse time, and the BH extractable energy (used to compute the NS critical mass)—scale together in a correlated manner.} 

\textcolor{black}{We further investigated the influence of jet collimation by adopting a median half-opening angle of \(\theta_j = 16^\circ \pm 10^\circ\), consistent with values derived from afterglow modeling of SGRBs \citep{2016ApJ...827..102T,2015ApJ...815..102F,2019cxo..prop.5648T}, specifically applied to GRB~051210. We assumed this opening angle remains constant both prior to and following the collapse. Under this assumption, the GM1 EOS continued to yield the statistically preferred fit; however, the inferred magnetar surface magnetic field reached values on the order of \(5 \times 10^{17}\,\mathrm{G}\), which may be considered unphysically high. To reconcile this, we incorporated the effects of radiation efficiency, adopting a plausible range of \(\epsilon \sim 0.05\% - 10\%\) \citep{nava2013afterglow,2019ApJDai}, alongside the jet collimation correction. This combined treatment does not significantly affect the model fits, with the GM1 EOS remaining favored. Although this analysis is based on a single GRB 051210, the consistency of our results across a plausible range of redshifts, jet opening angles, and radiative efficiencies suggests that these parameters may not significantly impact our conclusions for similar bursts. To further quantify potential systematic uncertainties related to redshift, efficiency, and jet collimation, in the forthcoming work we intend to extend our analysis to a larger sample of over 50 SGRBs with robust redshift measurements \citep[see][]{2022MNRAS.515.4890O}. }

  Our 3 EOS represent a limited subset; phase transitions or other physics may alter $M_{\rm crit}$ \citep{2015ApJ...798L..36C,2015ApJ...802...95R}. Numerical simulations (e.g., general relativistic hydrostatic equilibrium code, RNS \citep{1995ApJ...444..306S})  should test broader EOS in future work. While the GM1 EOS provides the best overall fit, for some GRBs the difference from TM1 (with $M_{\rm crit}=2.2\,M_\odot$) remains marginal, with $\Delta\langle\Delta P\rangle\sim10\%$.  This corresponds to  
$M_{\rm crit}\approx2.3\pm0.1\,M_\odot$ rather than definitively excluding either softer or stiffer equations of state (Figs.~\ref{fig:LC} and \ref{fig:080905AF}).  This uncertainty in $M_{\rm crit}$ scales inversely with the amount of data included: broader temporal resolution coverage yields smaller error bars.  In fact, improved data quality leads to tighter EOS constraints: restricting our analysis to the highest-quality light-curve subset shifts the central value to $M_{\rm crit}\approx2.4\,M_\odot$ and markedly reduces the uncertainty.

While numerical codes can also infer the critical mass of neutron stars and obtain appropriate EOS that satisfy observational data for both neutron stars and black holes, our approach emphasizes the theoretical framework due to its robustness and general applicability \citep{lattimer2001}. 
\textcolor{black}{Several earlier investigations have used SGRB signatures to probe the EOS, including tests based on quasi-periodic oscillations in the post-merger phase \citep{2023Natur.613..253C,2025ApJ...980..220H}, precursors \citep{2012PhRvL.108a1102T,2023ApJ...954L..29D}, and GRB duration \citep{2025PhRvD.111f3015P}.}

\textcolor{black}{Our work builds upon previous studies that employ X-ray plateau durations to constrain the neutron star equation of state \citep{2014MNRAS.441.2433R,lasky2014,2015ApJ...805...89L,2024A&A...692A..73G}, including more recent efforts that incorporate fallback accretion into the post-merger spin-down evolution \citep{2024A&A...692A..73G}. These investigations typically utilize plateau durations to estimate the collapse time and derive EOS constraints within dipolar spin down and critical mass of the neutron star framework, with some studies finding a preference for a softer EOS when accounting for fallback accretion in the post-merger phase \citep[e.g.,][]{2022ApJ...939...51M}.}

\textcolor{black}{In contrast, the present analysis focuses on a different aspect: we link the critical neutron star mass to black hole formation criteria and, in a novel step, directly apply the black hole mass formulation to evaluate the extractable rotational energy. This black-hole–centred treatment, which has not been applied in earlier plateau-based EOS studies, shifts the emphasis to quantifying the energy budget available from the remnant BH itself. Here we neglect fallback accretion in order to isolate the impact of higher-order multipolar spin-down terms together with the BH mass–energy relation, while noting that incorporating fallback physics might be an important extension in future work. This approach offers a complementary pathway for interpreting the plateau phase and deriving EOS constraints based on a different set of governing parameters.}

\textcolor{black}{This proof-of-concept confirms the feasibility of combining multipolar spin-down with BH mass–energy considerations; forthcoming studies will extend the analysis to additional EOS models, implement dedicated numerical simulations \citep{1995ApJ...444..306S}, and incorporate gravitational-wave emission effects \citep{2015PhRvD..91f4001T,2019JPhG...46k3002B,2022EPJWC.27401013M} and fallback accretion, to enable a direct comparison with existing post-merger models.}

\section{Conclusion}\label{sec:conclusion}

We have presented a novel framework to constrain the neutron-star equation of state (EOS) by linking short gamma-ray burst (SGRB) X-ray ``internal plateaus'' to supramassive magnetar collapse and subsequent Kerr-black-hole energy extraction. For 10 SGRBs with well-observed plateaus and steep decays, our analysis favors an intermediate-stiffness EOS (GM1, $M_{\rm TOV} = 2.39\,M_\odot$), minimizing discrepancies ($\Delta P = |P_{\rm NS} - P_{\rm BH}|$) between pre-collapse spin periods (from multipolar spin-down fits; \citealt{2024ApJ...974...89W,Yorgancioglu2025Dainotti}) and post-collapse Kerr BH constraints. Softer (TM1, $M_{\rm TOV}=2.2\,M_\odot$) and stiffer (NL3, $M_{\rm TOV}=2.8\,M_\odot$) extremes are disfavored. This result is robust against redshift uncertainties due to self-consistent bolometric scaling \citep{rowlinson2013} and aligns with the Dainotti correlation \citep{2008MNRAS.391L..79D,Dainotti:2022ked}, suggesting magnetar-powered plateaus as the dominant mechanism \citep{2006MNRAS.372L..19F,2007ApJ...665..599T,2007ApJ...670..565L,2010MNRAS.402..705L,2013MNRAS.431.1745G,2019ApJS..245....1T,2024ApJ...974...89W,Yorgancioglu2025Dainotti}. 

Our approach assumes: (1) plateaus originate from magnetar multipolar electromagnetic radiation rather than external shocks \citep{2015ApJ...805...13L}, (2) full extraction of Kerr black-hole rotational energy without significant accretion contributions \citep{2018MNRAS.475..266Z}, and (3) negligible gravitational-wave losses \citep{2015PhRvD..91f4001T}. {\color{black}We have further tested the robustness of these assumptions against uncertainties in redshift, jet opening angle, and radiative efficiency, finding that the preferred EOS (GM1) remains favored under reasonable variations.} Deviations from these assumptions, {\color{black}including possible fallback accretion, partial rather than full Kerr-BH energy extraction, or early gravitational-wave–dominated spin-down,} or EOS softening from phase transitions \citep{2020NatPh..16..907A}, could systematically bias $M_{\rm TOV}$. {\color{black}In addition, while the GM1 EOS (intermediate stiffness) minimizes $\Delta P$ for most bursts,  TM1 can be competitive at low redshifts or for alternative efficiency assumptions, indicating that the current constraint of $M_{\rm TOV}$ should be regarded as model-dependent.} While GM1 degeneracy persists for some bursts ($\Delta\langle\Delta P\rangle \sim 10\%$), suggesting $M_{\rm TOV} \approx 2.3 \pm 0.1\,M_\odot$, {\color{black}our results demonstrate that the combination of multipolar spin-down modeling with BH mass–energy constraints provides a complementary pathway to EOS determination, independent of purely dipolar spin-down or external-shock frameworks.} This proof-of-concept establishes SGRB plateaus as a viable dense-matter probe. Future work will leverage larger samples (e.g., \textit{Swift}-XRT/\textit{Einstein} Probe catalogues), extended EOS grids including hybrid/quark matter \citep{2024NatAs...8.1020M}, {\color{black}numerical RNS modeling across a broader parameter space, and incorporation of fallback accretion and gravitational-wave losses} \citep{1995ApJ...444..306S} to refine constraints and test phase transitions.

\bigskip

\acknowledgments
{\color{black} We thank the referee for their careful review and comments, which have helped to improve the technical clarity and methodological description of our analysis. The authors used an AI-based tool only to improve the grammar and readability of the manuscript.}
R. Moradi acknowledges support from the Academy of Sciences Beijing Natural Science Foundation (IS24021) and the Institute of High Energy Physics, Chinese(E32984U810).



\begin{thebibliography}{}
\expandafter\ifx\csname natexlab\endcsname\relax\def\natexlab#1{#1}\fi
\providecommand{\url}[1]{\href{#1}{#1}}

\bibitem[{{Abbott} {et~al.}(2019){Abbott}, {Abbott}, {Abbott}, {Acernese}, {Ackley}, {Adams}, {Adams}, {Addesso}, {Adhikari}, {Adya}, {Affeldt}, {Agarwal}, {Agathos}, {Agatsuma}, {Aggarwal}, {Aguiar}, {Aiello}, {Ain}, {Ajith}, {Allen}, {Allen}, {Allocca}, {Aloy}, {Altin}, {Amato}, {Ananyeva}, {Anderson}, {Anderson}, {Angelova}, {Antier}, {Appert}, {Arai}, {Araya}, {Areeda}, {Ar{\`e}ne}, {Arnaud}, {Arun}, {Ascenzi}, {Ashton}, {Ast}, {Aston}, {Astone}, {Atallah}, {Aubin}, {Aufmuth}, {Aulbert}, {AultONeal}, {Austin}, {Avila-Alvarez}, {Babak}, {Bacon}, {Badaracco}, {Bader}, {Bae}, {Baker}, {Baldaccini}, {Ballardin}, {Ballmer}, {Banagiri}, {Barayoga}, {Barclay}, {Barish}, {Barker}, {Barkett}, {Barnum}, {Barone}, {Barr}, {Barsotti}, {Barsuglia}, {Barta}, {Bartlett}, {Bartos}, {Bassiri}, {Basti}, {Batch}, {Bawaj}, {Bayley}, {Bazzan}, {B{\'e}csy}, {Beer}, {Bejger}, {Belahcene}, {Bell}, {Beniwal}, {Bensch}, {Berger}, {Bergmann}, {Bernuzzi}, {Bero}, {Berry}, {Bersanetti}, {Bertolini}, {Betzwieser}, {Bhandare},
  {Bilenko}, {Bilgili}, {Billingsley}, {Billman}, {Birch}, {Birney}, {Birnholtz}, {Biscans}, {Biscoveanu}, {Bisht}, {Bitossi}, {Bizouard}, {Blackburn}, {Blackman}, {Blair}, {Blair}, {Blair}, {Bloemen}, {Bock}, {Bode}, {Boer}, {Boetzel}, {Bogaert}, {Bohe}, {Bondu}, {Bonilla}, {Bonnand}, {Booker}, {Boom}, {Booth}, {Bork}, {Boschi}, {Bose}, {Bossie}, {Bossilkov}, {Bosveld}, {Bouffanais}, {Bozzi}, {Bradaschia}, {Brady}, {Bramley}, {Branchesi}, {Brau}, {Briant}, {Brighenti}, {Brillet}, {Brinkmann}, {Brisson}, {Brockill}, {Brooks}, {Brown}, {Brunett}, {Buchanan}, {Buikema}, {Bulik}, {Bulten}, {Buonanno}, {Buskulic}, {Buy}, {Byer}, {Cabero}, {Cadonati}, {Cagnoli}, {Cahillane}, {Bustillo}, {Callister}, {Calloni}, {Camp}, {Canepa}, {Canizares}, {Cannon}, {Cao}, {Cao}, {Capano}, {Capocasa}, {Carbognani}, {Caride}, {Carney}, {Carullo}, {Diaz}, {Casentini}, {Caudill}, {Cavagli{\`a}}, {Cavalier}, {Cavalieri}, {Cella}, {Cepeda}, {Cerd{\'a}-Dur{\'a}n}, {Cerretani}, {Cesarini}, {Chaibi}, {Chamberlin}, {Chan}, {Chao},
  {Charlton}, {Chase}, {Chassande-Mottin}, {Chatterjee}, {Chatziioannou}, {Cheeseboro}, {Chen}, {Chen}, {Chen}, {Cheng}, {Chia}, {Chincarini}, {Chiummo}, {Chmiel}, {Cho}, {Cho}, {Chow}, {Christensen}, {Chu}, {Chua}, {Chua}, {Chung}, {Chung}, {Ciani}, {Ciobanu}, {Ciolfi}, {Cipriano}, {Cirelli}, {Cirone}, {Clara}, {Clark}, {Clearwater}, {Cleva}, {Cocchieri}, {Coccia}, {Cohadon}, {Cohen}, {Colla}, {Collette}, {Collins}, {Cominsky}, {Constancio}, {Conti}, {Cooper}, {Corban}, {Corbitt}, {Cordero-Carri{\'o}n}, {Corley}, {Cornish}, {Corsi}, {Cortese}, {Costa}, {Cotesta}, {Coughlin}, {Coughlin}, {Coulon}, {Countryman}, {Couvares}, {Covas}, {Cowan}, {Coward}, {Cowart}, {Coyne}, {Coyne}, {Creighton}, {Creighton}, {Cripe}, {Crowder}, {Cullen}, {Cumming}, {Cunningham}, {Cuoco}, {Canton}, {D{\'a}lya}, {Danilishin}, {D'Antonio}, {Danzmann}, {Dasgupta}, {Costa}, {Dattilo}, {Dave}, {Davier}, {Davis}, {Daw}, {Day}, {DeBra}, {Deenadayalan}, {Degallaix}, {De Laurentis}, {Del{\'e}glise}, {Del Pozzo}, {Demos}, {Denker}, {Dent},
  {De Pietri}, {Derby}, {Dergachev}, {De Rosa}, {De Rossi}, {DeSalvo}, {de Varona}, {Dhurandhar}, {D{\'\i}az}, {Dietrich}, {Di Fiore}, {Di Giovanni}, {Di Girolamo}, {Di Lieto}, {Ding}, {Di Pace}, {Di Palma}, {Di Renzo}, {Dmitriev}, {Doctor}, {Dolique}, {Donovan}, {Dooley}, {Doravari}, {Dorrington}, {{\'A}lvarez}, {Downes}, {Drago}, {Dreissigacker}, {Driggers}, {Du}, {Dudi}, {Dupej}, {Dwyer}, {Easter}, {Edo}, {Edwards}, {Effler}, {Eggenstein}, {Ehrens}, {Eichholz}, {Eikenberry}, {Eisenmann}, {Eisenstein}, {Essick}, {Estelles}, {Estevez}, {Etienne}, {Etzel}, {Evans}, {Evans}, {Fafone}, {Fair}, {Fairhurst}, {Fan}, {Farinon}, {Farr}, {Farr}, {Fauchon-Jones}, {Favata}, {Fays}, {Fee}, {Fehrmann}, {Feicht}, {Fejer}, {Feng}, {Fernandez-Galiana}, {Ferrante}, {Ferreira}, {Ferrini}, {Fidecaro}, {Fiori}, {Fiorucci}, {Fishbach}, {Fisher}, {Fishner}, {Fitz-Axen}, {Flaminio}, {Fletcher}, {Fong}, {Font}, {Forsyth}, {Forsyth}, {Fournier}, {Frasca}, {Frasconi}, {Frei}, {Freise}, {Frey}, {Frey}, {Fritschel}, {Frolov}, {Fulda},
  {Fyffe}, {Gabbard}, {Gadre}, {Gaebel}, {Gair}, {Gammaitoni}, {Ganija}, {Gaonkar}, {Garcia}, {Garc{\'\i}a-Quir{\'o}s}, {Garufi}, {Gateley}, {Gaudio}, {Gaur}, {Gayathri}, {Gemme}, {Genin}, {Gennai}, {George}, {George}, {Gergely}, {Germain}, {Ghonge}, {Ghosh}, {Ghosh}, {Ghosh}, {Giacomazzo}, {Giaime}, {Giardina}, {Giazotto}, {Gill}, {Giordano}, {Glover}, {Goetz}, {Goetz}, {Goncharov}, {Gonz{\'a}lez}, {Castro}, {Gopakumar}, {Gorodetsky}, {Gossan}, {Gosselin}, {Gouaty}, {Grado}, {Graef}, {Granata}, {Grant}, {Gras}, {Gray}, {Greco}, {Green}, {Green}, {Gretarsson}, {Groot}, {Grote}, {Grunewald}, {Gruning}, {Guidi}, {Gulati}, {Guo}, {Gupta}, {Gupta}, {Gushwa}, {Gustafson}, {Gustafson}, {Halim}, {Hall}, {Hall}, {Hamilton}, {Hamilton}, {Hammond}, {Haney}, {Hanke}, {Hanks}, {Hanna}, {Hannam}, {Hannuksela}, {Hanson}, {Hardwick}, {Harms}, {Harry}, {Harry}, {Hart}, {Haster}, {Haughian}, {Healy}, {Heidmann}, {Heintze}, {Heitmann}, {Hello}, {Hemming}, {Hendry}, {Heng}, {Hennig}, {Heptonstall}, {Hernandez}, {Heurs}, {Hild},
  {Hinderer}, {Hoak}, {Hochheim}, {Hofman}, {Holland}, {Holt}, {Holz}, {Hopkins}, {Horst}, {Hough}, {Houston}, {Howell}, {Hreibi}, {Huerta}, {Huet}, {Hughey}, {Hulko}, {Husa}, {Huttner}, {Huynh-Dinh}, {Iess}, {Indik}, {Ingram}, {Inta}, {Intini}, {Isa}, {Isac}, {Isi}, {Iyer}, {Izumi}, {Jacqmin}, {Jani}, {Jaranowski}, {Johnson}, {Johnson}, {Jones}, {Jones}, {Jonker}, {Ju}, {Junker}, {Kalaghatgi}, {Kalogera}, {Kamai}, {Kandhasamy}, {Kang}, {Kanner}, {Kapadia}, {Karki}, {Karvinen}, {Kasprzack}, {Kastaun}, {Katolik}, {Katsanevas}, {Katsavounidis}, {Katzman}, {Kaufer}, {Kawabe}, {Keerthana}, {K{\'e}f{\'e}lian}, {Keitel}, {Kemball}, {Kennedy}, {Key}, {Khalili}, {Khamesra}, {Khan}, {Khan}, {Khan}, {Khan}, {Khazanov}, {Kijbunchoo}, {Kim}, {Kim}, {Kim}, {Kim}, {Kim}, {Kim}, {King}, {King}, {Kinley-Hanlon}, {Kirchhoff}, {Kissel}, {Kleybolte}, {Klimenko}, {Knowles}, {Koch}, {Koehlenbeck}, {Koley}, {Kondrashov}, {Kontos}, {Korobko}, {Korth}, {Kowalska}, {Kozak}, {Kr{\"a}mer}, {Kringel}, {Krishnan}, {Kr{\'o}lak}, {Kuehn},
  {Kumar}, {Kumar}, {Kumar}, {Kuo}, {Kutynia}, {Kwang}, {Lackey}, {Lai}, {Landry}, {Landry}, {Lang}, {Lange}, {Lantz}, {Lanza}, {Lartaux-Vollard}, {Lasky}, {Laxen}, {Lazzarini}, {Lazzaro}, {Leaci}, {Leavey}, {Lee}, {Lee}, {Lee}, {Lee}, {Lee}, {Lehmann}, {Lenon}, {Leonardi}, {Leroy}, {Letendre}, {Levin}, {Li}, {Li}, {Li}, {Linker}, {Littenberg}, {Liu}, {Liu}, {Lo}, {Lockerbie}, {London}, {Longo}, {Lorenzini}, {Loriette}, {Lormand}, {Losurdo}, {Lough}, {Lousto}, {Lovelace}, {L{\"u}ck}, {Lumaca}, {Lundgren}, {Lynch}, {Ma}, {Macas}, {Macfoy}, {Machenschalk}, {MacInnis}, {Macleod}, {Hernandez}, {Maga{\~n}a-Sandoval}, {Zertuche}, {Magee}, {Majorana}, {Maksimovic}, {Man}, {Mandic}, {Mangano}, {Mansell}, {Manske}, {Mantovani}, {Marchesoni}, {Marion}, {M{\'a}rka}, {M{\'a}rka}, {Markakis}, {Markosyan}, {Markowitz}, {Maros}, {Marquina}, {Martelli}, {Martellini}, {Martin}, {Martin}, {Martynov}, {Mason}, {Massera}, {Masserot}, {Massinger}, {Masso-Reid}, {Mastrogiovanni}, {Matas}, {Matichard}, {Matone}, {Mavalvala},
  {Mazumder}, {McCann}, {McCarthy}, {McClelland}, {McCormick}, {McCuller}, {McGuire}, {McIver}, {McManus}, {McRae}, {McWilliams}, {Meacher}, {Meadors}, {Mehmet}, {Meidam}, {Mejuto-Villa}, {Melatos}, {Mendell}, {Mendoza-Gandara}, {Mercer}, {Mereni}, {Merilh}, {Merzougui}, {Meshkov}, {Messenger}, {Messick}, {Metzdorff}, {Meyers}, {Miao}, {Michel}, {Middleton}, {Mikhailov}, {Milano}, {Miller}, {Miller}, {Miller}, {Miller}, {Millhouse}, {Mills}, {Milovich-Goff}, {Minazzoli}, {Minenkov}, {Ming}, {Mishra}, {Mitra}, {Mitrofanov}, {Mitselmakher}, {Mittleman}, {Moffa}, {Mogushi}, {Mohan}, {Mohapatra}, {Montani}, {Moore}, {Moraru}, {Moreno}, {Morisaki}, {Mours}, {Mow-Lowry}, {Mueller}, {Muir}, {Mukherjee}, {Mukherjee}, {Mukherjee}, {Mukund}, {Mullavey}, {Munch}, {Mu{\~n}iz}, {Muratore}, {Murray}, {Nagar}, {Napier}, {Nardecchia}, {Naticchioni}, {Nayak}, {Neilson}, {Nelemans}, {Nelson}, {Nery}, {Neunzert}, {Nevin}, {Newport}, {Ng}, {Ng}, {Nguyen}, {Nguyen}, {Nichols}, {Nielsen}, {Nissanke}, {Nitz}, {Nocera}, {Nolting},
  {North}, {Nuttall}, {Obergaulinger}, {Oberling}, {O'Brien}, {O'Dea}, {Ogin}, {Oh}, {Oh}, {Ohme}, {Ohta}, {Okada}, {Oliver}, {Oppermann}, {Oram}, {O'Reilly}, {Ormiston}, {Ortega}, {O'Shaughnessy}, {Ossokine}, {Ottaway}, {Overmier}, {Owen}, {Pace}, {Pagano}, {Page}, {Page}, {Pai}, {Pai}, {Palamos}, {Palashov}, {Palomba}, {Pal-Singh}, {Pan}, {Pan}, {Pang}, {Pang}, {Pankow}, {Pannarale}, {Pant}, {Paoletti}, {Paoli}, {Papa}, {Parida}, {Parker}, {Pascucci}, {Pasqualetti}, {Passaquieti}, {Passuello}, {Patil}, {Patricelli}, {Pearlstone}, {Pedersen}, {Pedraza}, {Pedurand}, {Pekowsky}, {Pele}, {Penn}, {Perez}, {Perreca}, {Perri}, {Pfeiffer}, {Phelps}, {Phukon}, {Piccinni}, {Pichot}, {Piergiovanni}, {Pierro}, {Pillant}, {Pinard}, {Pinto}, {Pirello}, {Pitkin}, {Poggiani}, {Popolizio}, {Porter}, {Possenti}, {Post}, {Powell}, {Prasad}, {Pratt}, {Pratten}, {Predoi}, {Prestegard}, {Principe}, {Privitera}, {Prodi}, {Prokhorov}, {Puncken}, {Punturo}, {Puppo}, {P{\"u}rrer}, {Qi}, {Quetschke}, {Quintero}, {Quitzow-James},
  {Raab}, {Rabeling}, {Radkins}, {Raffai}, {Raja}, {Rajan}, {Rajbhandari}, {Rakhmanov}, {Ramirez}, {Ramos-Buades}, {Rana}, {Rapagnani}, {Raymond}, {Razzano}, {Read}, {Regimbau}, {Rei}, {Reid}, {Reitze}, {Ren}, {Ricci}, {Ricker}, {Riemenschneider}, {Riles}, {Rizzo}, {Robertson}, {Robie}, {Robinet}, {Robson}, {Rocchi}, {Rolland}, {Rollins}, {Roma}, {Romano}, {Romel}, {Romie}, {Rosi{\'n}ska}, {Ross}, {Rowan}, {R{\"u}diger}, {Ruggi}, {Rutins}, {Ryan}, {Sachdev}, {Sadecki}, {Sakellariadou}, {Salconi}, {Saleem}, {Salemi}, {Samajdar}, {Sammut}, {Sampson}, {Sanchez}, {Sanchez}, {Sanchis-Gual}, {Sandberg}, {Sanders}, {Sarin}, {Sassolas}, {Sathyaprakash}, {Saulson}, {Sauter}, {Savage}, {Sawadsky}, {Schale}, {Scheel}, {Scheuer}, {Schmidt}, {Schnabel}, {Schofield}, {Sch{\"o}nbeck}, {Schreiber}, {Schuette}, {Schulte}, {Schutz}, {Schwalbe}, {Scott}, {Scott}, {Seidel}, {Sellers}, {Sengupta}, {Sentenac}, {Sequino}, {Sergeev}, {Setyawati}, {Shaddock}, {Shaffer}, {Shah}, {Shahriar}, {Shaner}, {Shao}, {Shapiro}, {Shawhan},
  {Shen}, {Shoemaker}, {Shoemaker}, {Siellez}, {Siemens}, {Sieniawska}, {Sigg}, {Silva}, {Singer}, {Singh}, {Singhal}, {Sintes}, {Slagmolen}, {Slaven-Blair}, {Smith}, {Smith}, {Smith}, {Somala}, {Son}, {Sorazu}, {Sorrentino}, {Souradeep}, {Spencer}, {Srivastava}, {Staats}, {Steinke}, {Steinlechner}, {Steinlechner}, {Steinmeyer}, {Steltner}, {Stevenson}, {Stocks}, {Stone}, {Stops}, {Strain}, {Stratta}, {Strigin}, {Strunk}, {Sturani}, {Stuver}, {Summerscales}, {Sun}, {Sunil}, {Suresh}, {Sutton}, {Swinkels}, {Szczepa{\'n}czyk}, {Tacca}, {Tait}, {Talbot}, {Talukder}, {Tanner}, {T{\'a}pai}, {Taracchini}, {Tasson}, {Taylor}, {Taylor}, {Tewari}, {Theeg}, {Thies}, {Thomas}, {Thomas}, {Thomas}, {Thorne}, {Thrane}, {Tiwari}, {Tiwari}, {Tokmakov}, {Toland}, {Tonelli}, {Tornasi}, {Torres-Forn{\'e}}, {Torrie}, {T{\"o}yr{\"a}}, {Travasso}, {Traylor}, {Trinastic}, {Tringali}, {Trozzo}, {Tsang}, {Tse}, {Tso}, {Tsuna}, {Tsukada}, {Tuyenbayev}, {Ueno}, {Ugolini}, {Urban}, {Usman}, {Vahlbruch}, {Vajente}, {Valdes}, {van Bakel},
  {van Beuzekom}, {van den Brand}, {Van Den Broeck}, {Vander-Hyde}, {van der Schaaf}, {van Heijningen}, {van Veggel}, {Vardaro}, {Varma}, {Vass}, {Vas{\'u}th}, {Vecchio}, {Vedovato}, {Veitch}, {Veitch}, {Venkateswara}, {Venugopalan}, {Verkindt}, {Vetrano}, {Vicer{\'e}}, {Viets}, {Vinciguerra}, {Vine}, {Vinet}, {Vitale}, {Vo}, {Vocca}, {Vorvick}, {Vyatchanin}, {Wade}, {Wade}, {Wade}, {Walet}, {Walker}, {Wallace}, {Walsh}, {Wang}, {Wang}, {Wang}, {Wang}, {Wang}, {Ward}, {Warner}, {Was}, {Watchi}, {Weaver}, {Wei}, {Weinert}, {Weinstein}, {Weiss}, {Wellmann}, {Wen}, {Wessel}, {We{\ss}els}, {Westerweck}, {Wette}, {Whelan}, {Whiting}, {Whittle}, {Wilken}, {Williams}, {Williams}, {Williamson}, {Willis}, {Willke}, {Wimmer}, {Winkler}, {Wipf}, {Wittel}, {Woan}, {Woehler}, {Wofford}, {Wong}, {Worden}, {Wright}, {Wu}, {Wysocki}, {Xiao}, {Yam}, {Yamamoto}, {Yancey}, {Yang}, {Yap}, {Yazback}, {Yu}, {Yu}, {Yvert}, {Zadro{\.z}ny}, {Zanolin}, {Zelenova}, {Zendri}, {Zevin}, {Zhang}, {Zhang}, {Zhang}, {Zhang}, {Zhang}, {Zhao},
  {Zhou}, {Zhou}, {Zhu}, {Zhu}, {Zimmerman}, {Zlochower}, {Zucker}, {Zweizig}, {LIGO Scientific Collaboration}, \& {Virgo Collaboration}}]{2019PhRvX...9a1001A}
{Abbott}, B.~P., {Abbott}, R., {Abbott}, T.~D., {et~al.} 2019, Physical Review X, 9, 011001

\bibitem[{{Akaike}(1974)}]{1974ITAC...19..716A}
{Akaike}, H. 1974, IEEE Transactions on Automatic Control, 19, 716

\bibitem[{{Annala} {et~al.}(2020){Annala}, {Gorda}, {Kurkela}, {N{\"a}ttil{\"a}}, \& {Vuorinen}}]{2020NatPh..16..907A}
{Annala}, E., {Gorda}, T., {Kurkela}, A., {N{\"a}ttil{\"a}}, J., \& {Vuorinen}, A. 2020, Nature Physics, 16, 907

\bibitem[{{Annala} {et~al.}(2018){Annala}, {Gorda}, {Kurkela}, \& {Vuorinen}}]{2018PhRvL.120q2703A}
{Annala}, E., {Gorda}, T., {Kurkela}, A., \& {Vuorinen}, A. 2018, \prl, 120, 172703

\bibitem[{{Band} {et~al.}(1993){Band}, {Matteson}, {Ford}, {Schaefer}, {Palmer}, {Teegarden}, {Cline}, {Briggs}, {Paciesas}, {Pendleton}, {Fishman}, {Kouveliotou}, {Meegan}, {Wilson}, \& {Lestrade}}]{1993ApJ...413..281B}
{Band}, D., {Matteson}, J., {Ford}, L., {et~al.} 1993, \apj, 413, 281

\bibitem[{Barrera {et~al.}(1985)Barrera, Estevez, \& Giraldo}]{barrera1985}
Barrera, R.~G., Estevez, G.~A., \& Giraldo, J. 1985, European Journal of Physics, 6, 287

\bibitem[{{Bauswein} \& {Stergioulas}(2019)}]{2019JPhG...46k3002B}
{Bauswein}, A., \& {Stergioulas}, N. 2019, Journal of Physics G Nuclear Physics, 46, 113002

\bibitem[{{Beniamini} {et~al.}(2020){Beniamini}, {Duque}, {Daigne}, \& {Mochkovitch}}]{2020MNRAS.492.2847B}
{Beniamini}, P., {Duque}, R., {Daigne}, F., \& {Mochkovitch}, R. 2020, \mnras, 492, 2847

\bibitem[{{Beniamini} \& {Lu}(2021)}]{beniamini2021}
{Beniamini}, P., \& {Lu}, W. 2021, \apj, 920, 109

\bibitem[{{Berger}(2014)}]{2014ARA&A..52...43B}
{Berger}, E. 2014, \araa, 52, 43

\bibitem[{{Blandford} \& {Znajek}(1977)}]{1977MNRAS.179..433B}
{Blandford}, R.~D., \& {Znajek}, R.~L. 1977, \mnras, 179, 433

\bibitem[{{Bloom} {et~al.}(2001){Bloom}, {Frail}, \& {Sari}}]{2001AJ....121.2879B}
{Bloom}, J.~S., {Frail}, D.~A., \& {Sari}, R. 2001, \aj, 121, 2879

\bibitem[{Burnham \& Anderson(2004)}]{doi:10.1177/0049124104268644}
Burnham, K.~P., \& Anderson, D.~R. 2004, Sociological Methods \& Research, 33, 261.
\newblock \url{https://doi.org/10.1177/0049124104268644}

\bibitem[{{Cannizzo} {et~al.}(2011){Cannizzo}, {Troja}, \& {Gehrels}}]{2011ApJ...734...35C}
{Cannizzo}, J.~K., {Troja}, E., \& {Gehrels}, N. 2011, \apj, 734, 35

\bibitem[{Casella \& Berger(2002)}]{casella2002statistical}
Casella, G., \& Berger, R.~L. 2002, Statistical Inference (Duxbury Press)

\bibitem[{{Chandra} \& {Frail}(2012)}]{2012ApJ...746..156C}
{Chandra}, P., \& {Frail}, D.~A. 2012, \apj, 746, 156

\bibitem[{{Chirenti} {et~al.}(2023){Chirenti}, {Dichiara}, {Lien}, {Miller}, \& {Preece}}]{2023Natur.613..253C}
{Chirenti}, C., {Dichiara}, S., {Lien}, A., {Miller}, M.~C., \& {Preece}, R. 2023, \nat, 613, 253

\bibitem[{{Christodoulou}(1970)}]{1970PhRvL..25.1596C}
{Christodoulou}, D. 1970, Physical Review Letters, 25, 1596

\bibitem[{{Christodoulou} \& {Ruffini}(1971)}]{1971PhRvD...4.3552C}
{Christodoulou}, D., \& {Ruffini}, R. 1971, \prd, 4, 3552

\bibitem[{{Ciolfi} \& {Siegel}(2015)}]{2015ApJ...798L..36C}
{Ciolfi}, R., \& {Siegel}, D.~M. 2015, \apjl, 798, L36

\bibitem[{{Cipolletta} {et~al.}(2015){Cipolletta}, {Cherubini}, {Filippi}, {Rueda}, \& {Ruffini}}]{2015PhRvD..92b3007C}
{Cipolletta}, F., {Cherubini}, C., {Filippi}, S., {Rueda}, J.~A., \& {Ruffini}, R. 2015, \prd, 92, 023007

\bibitem[{{Dainotti} {et~al.}(2008){Dainotti}, {Cardone}, \& {Capozziello}}]{2008MNRAS.391L..79D}
{Dainotti}, M.~G., {Cardone}, V.~F., \& {Capozziello}, S. 2008, \mnras, 391, L79

\bibitem[{{Dainotti} {et~al.}(2011{\natexlab{a}}){Dainotti}, {Fabrizio Cardone}, {Capozziello}, {Ostrowski}, \& {Willingale}}]{2011ApJ...730..135D}
{Dainotti}, M.~G., {Fabrizio Cardone}, V., {Capozziello}, S., {Ostrowski}, M., \& {Willingale}, R. 2011{\natexlab{a}}, \apj, 730, 135

\bibitem[{Dainotti {et~al.}(2023)Dainotti, Lenart, Chraya, Sarracino, Nagataki, Fraija, Capozziello, \& Bogdan}]{Dainotti:2022ked}
Dainotti, M.~G., Lenart, A.~L., Chraya, A., {et~al.} 2023, Mon. Not. Roy. Astron. Soc., 518, 2201

\bibitem[{{Dainotti} {et~al.}(2011{\natexlab{b}}){Dainotti}, {Ostrowski}, \& {Willingale}}]{2011MNRAS.418.2202D}
{Dainotti}, M.~G., {Ostrowski}, M., \& {Willingale}, R. 2011{\natexlab{b}}, \mnras, 418, 2202

\bibitem[{{Dainotti} {et~al.}(2013){Dainotti}, {Petrosian}, {Singal}, \& {Ostrowski}}]{2013ApJ...774..157D}
{Dainotti}, M.~G., {Petrosian}, V., {Singal}, J., \& {Ostrowski}, M. 2013, \apj, 774, 157

\bibitem[{{Dainotti} {et~al.}(2010){Dainotti}, {Willingale}, {Capozziello}, {Fabrizio Cardone}, \& {Ostrowski}}]{2010ApJ...722L.215D}
{Dainotti}, M.~G., {Willingale}, R., {Capozziello}, S., {Fabrizio Cardone}, V., \& {Ostrowski}, M. 2010, \apjl, 722, L215

\bibitem[{{Dall'Osso} {et~al.}(2011){Dall'Osso}, {Stratta}, {Guetta}, {Covino}, {De Cesare}, \& {Stella}}]{2011A&A...526A.121D}
{Dall'Osso}, S., {Stratta}, G., {Guetta}, D., {et~al.} 2011, \aap, 526, A121

\bibitem[{{Dichiara} {et~al.}(2023){Dichiara}, {Tsang}, {Troja}, {Neill}, {Norris}, \& {Yang}}]{2023ApJ...954L..29D}
{Dichiara}, S., {Tsang}, D., {Troja}, E., {et~al.} 2023, \apjl, 954, L29

\bibitem[{{Eichler} {et~al.}(1989){Eichler}, {Livio}, {Piran}, \& {Schramm}}]{eichler1989}
{Eichler}, D., {Livio}, M., {Piran}, T., \& {Schramm}, D.~N. 1989, \nat, 340, 126

\bibitem[{{Evans} {et~al.}(2007{\natexlab{a}}){Evans}, {Beardmore}, {Page}, {Tyler}, {Osborne}, {Goad}, {O'Brien}, {Vetere}, {Racusin}, {Morris}, {Burrows}, {Capalbi}, {Perri}, {Gehrels}, \& {Romano}}]{2007A&A...469..379E}
{Evans}, P.~A., {Beardmore}, A.~P., {Page}, K.~L., {et~al.} 2007{\natexlab{a}}, \aap, 469, 379

\bibitem[{{Evans} {et~al.}(2007{\natexlab{b}}){Evans}, {Beardmore}, {Page}, {Tyler}, {Osborne}, {Goad}, {O'Brien}, {Vetere}, {Racusin}, {Morris}, {Burrows}, {Capalbi}, {Perri}, {Gehrels}, \& {Romano}}]{Evans2007}
---. 2007{\natexlab{b}}, \aap, 469, 379

\bibitem[{{Evans} {et~al.}(2009){Evans}, {Beardmore}, {Page}, {Osborne}, {O'Brien}, {Willingale}, {Starling}, {Burrows}, {Godet}, {Vetere}, {Racusin}, {Goad}, {Wiersema}, {Angelini}, {Capalbi}, {Chincarini}, {Gehrels}, {Kennea}, {Margutti}, {Morris}, {Mountford}, {Pagani}, {Perri}, {Romano}, \& {Tanvir}}]{Evans2009}
---. 2009, \mnras, 397, 1177

\bibitem[{{Falcke} \& {Rezzolla}(2014)}]{2014A&A...562A.137F}
{Falcke}, H., \& {Rezzolla}, L. 2014, \aap, 562, A137

\bibitem[{{Fan} \& {Xu}(2006)}]{2006MNRAS.372L..19F}
{Fan}, Y.-Z., \& {Xu}, D. 2006, \mnras, 372, L19

\bibitem[{{Fong} {et~al.}(2015){Fong}, {Berger}, {Margutti}, \& {Zauderer}}]{2015ApJ...815..102F}
{Fong}, W., {Berger}, E., {Margutti}, R., \& {Zauderer}, B.~A. 2015, \apj, 815, 102

\bibitem[{{Gompertz} {et~al.}(2013){Gompertz}, {O'Brien}, {Wynn}, \& {Rowlinson}}]{2013MNRAS.431.1745G}
{Gompertz}, B.~P., {O'Brien}, P.~T., {Wynn}, G.~A., \& {Rowlinson}, A. 2013, \mnras, 431, 1745

\bibitem[{{Gruber} {et~al.}(2014){Gruber}, {Goldstein}, {Weller von Ahlefeld}, {Narayana Bhat}, {Bissaldi}, {Briggs}, {Byrne}, {Cleveland}, {Connaughton}, {Diehl}, {Fishman}, {Fitzpatrick}, {Foley}, {Gibby}, {Giles}, {Greiner}, {Guiriec}, {van der Horst}, {von Kienlin}, {Kouveliotou}, {Layden}, {Lin}, {Meegan}, {McGlynn}, {Paciesas}, {Pelassa}, {Preece}, {Rau}, {Wilson-Hodge}, {Xiong}, {Younes}, \& {Yu}}]{2014ApJS..211...12G}
{Gruber}, D., {Goldstein}, A., {Weller von Ahlefeld}, V., {et~al.} 2014, \apjs, 211, 12

\bibitem[{{Guglielmi} {et~al.}(2024){Guglielmi}, {Stratta}, {Dall'Osso}, {Singh}, {Brusa}, \& {Perna}}]{2024A&A...692A..73G}
{Guglielmi}, L., {Stratta}, G., {Dall'Osso}, S., {et~al.} 2024, \aap, 692, A73

\bibitem[{{Hawking}(1971)}]{1971PhRvL..26.1344H}
{Hawking}, S.~W. 1971, Physical Review Letters, 26, 1344

\bibitem[{{Huang} {et~al.}(2025){Huang}, {L{\"u}}, \& {Liang}}]{2025ApJ...980..220H}
{Huang}, J.-X., {L{\"u}}, H.-J., \& {Liang}, E.-W. 2025, \apj, 980, 220

\bibitem[{{Jackson}(1998)}]{1998clel.book.....J}
{Jackson}, J.~D. 1998, {Classical Electrodynamics, 3rd Edition}, 832

\bibitem[{{Kisaka} \& {Ioka}(2015)}]{2015ApJ...804L..16K}
{Kisaka}, S., \& {Ioka}, K. 2015, \apjl, 804, L16

\bibitem[{{Komissarov}(2004)}]{2004MNRAS.350..427K}
{Komissarov}, S.~S. 2004, \mnras, 350, 427

\bibitem[{{Komissarov} \& {Barkov}(2009)}]{2009MNRAS.397.1153K}
{Komissarov}, S.~S., \& {Barkov}, M.~V. 2009, \mnras, 397, 1153

\bibitem[{{Lasky} {et~al.}(2014){Lasky}, {Haskell}, {Ravi}, {Howell}, \& {Coward}}]{lasky2014}
{Lasky}, P.~D., {Haskell}, B., {Ravi}, V., {Howell}, E.~J., \& {Coward}, D.~M. 2014, \prd, 89, 047302

\bibitem[{{Lattimer} \& {Prakash}(2001)}]{lattimer2001}
{Lattimer}, J.~M., \& {Prakash}, M. 2001, \apj, 550, 426

\bibitem[{{Li} {et~al.}(2018){Li}, {Wang}, {Shao}, {Wu}, {Huang}, {Zhang}, {Ryde}, \& {Yu}}]{2018ApJS..234...26L}
{Li}, L., {Wang}, Y., {Shao}, L., {et~al.} 2018, \apjs, 234, 26

\bibitem[{{Li} {et~al.}(2015){Li}, {Wu}, {Huang}, {Wang}, {Tang}, {Liang}, {Zhang}, {Wang}, {Geng}, {Liang}, {Wei}, {Zhang}, \& {Ryde}}]{2015ApJ...805...13L}
{Li}, L., {Wu}, X.-F., {Huang}, Y.-F., {et~al.} 2015, \apj, 805, 13

\bibitem[{{Liang} {et~al.}(2007){Liang}, {Zhang}, \& {Zhang}}]{2007ApJ...670..565L}
{Liang}, E.-W., {Zhang}, B.-B., \& {Zhang}, B. 2007, \apj, 670, 565

\bibitem[{{Louren{\c{c}}o} {et~al.}(2020){Louren{\c{c}}o}, {Bhuyan}, {Lenzi}, {Dutra}, {Gonzalez-Boquera}, {Centelles}, \& {Vi{\~n}as}}]{2020PhLB..80335306L}
{Louren{\c{c}}o}, O., {Bhuyan}, M., {Lenzi}, C.~H., {et~al.} 2020, Physics Letters B, 803, 135306

\bibitem[{{L{\"u}} {et~al.}(2015){L{\"u}}, {Zhang}, {Lei}, {Li}, \& {Lasky}}]{2015ApJ...805...89L}
{L{\"u}}, H.-J., {Zhang}, B., {Lei}, W.-H., {Li}, Y., \& {Lasky}, P.~D. 2015, \apj, 805, 89

\bibitem[{{Lyons} {et~al.}(2010){Lyons}, {O'Brien}, {Zhang}, {Willingale}, {Troja}, \& {Starling}}]{2010MNRAS.402..705L}
{Lyons}, N., {O'Brien}, P.~T., {Zhang}, B., {et~al.} 2010, \mnras, 402, 705

\bibitem[{{Margalit} {et~al.}(2022){Margalit}, {Jermyn}, {Metzger}, {Roberts}, \& {Quataert}}]{2022ApJ...939...51M}
{Margalit}, B., {Jermyn}, A.~S., {Metzger}, B.~D., {Roberts}, L.~F., \& {Quataert}, E. 2022, \apj, 939, 51

\bibitem[{{Margalit} \& {Metzger}(2019)}]{2019ApJ...880L..15M}
{Margalit}, B., \& {Metzger}, B.~D. 2019, \apjl, 880, L15

\bibitem[{{Margalit} {et~al.}(2015){Margalit}, {Metzger}, \& {Beloborodov}}]{2015PhRvL.115q1101M}
{Margalit}, B., {Metzger}, B.~D., \& {Beloborodov}, A.~M. 2015, \prl, 115, 171101

\bibitem[{{Marino} {et~al.}(2024){Marino}, {Dehman}, {Kovlakas}, {Rea}, {Pons}, \& {Vigan{\`o}}}]{2024NatAs...8.1020M}
{Marino}, A., {Dehman}, C., {Kovlakas}, K., {et~al.} 2024, Nature Astronomy, 8, 1020

\bibitem[{{Mathews} {et~al.}(2022){Mathews}, {Kedia}, {Kim}, \& {Suh}}]{2022EPJWC.27401013M}
{Mathews}, G.~J., {Kedia}, A., {Kim}, H.~I., \& {Suh}, I.-S. 2022, in European Physical Journal Web of Conferences, Vol. 274, European Physical Journal Web of Conferences (EDP), 01013

\bibitem[{{Miller} {et~al.}(2019){Miller}, {Lamb}, {Dittmann}, {Bogdanov}, {Arzoumanian}, {Gendreau}, {Guillot}, {Harding}, {Ho}, {Lattimer}, {Ludlam}, {Mahmoodifar}, {Morsink}, {Ray}, {Strohmayer}, {Wood}, {Enoto}, {Foster}, {Okajima}, {Prigozhin}, \& {Soong}}]{2019ApJ...887L..24M}
{Miller}, M.~C., {Lamb}, F.~K., {Dittmann}, A.~J., {et~al.} 2019, \apjl, 887, L24

\bibitem[{Mészáros \& Rees(1993)}]{meszarosrees93}
Mészáros, P., \& Rees, M.~J. 1993, The Astrophysical Journal, 405, 278.
\newblock \url{https://ui.adsabs.harvard.edu/abs/1993ApJ...405..278M}

\bibitem[{Mészáros \& Rees(1997)}]{meszarosrees97}
---. 1997, The Astrophysical Journal, 476, 232.
\newblock \url{https://ui.adsabs.harvard.edu/abs/1997ApJ...476..232M}

\bibitem[{{Narayan} {et~al.}(1992){Narayan}, {Paczynski}, \& {Piran}}]{narayan1992}
{Narayan}, R., {Paczynski}, B., \& {Piran}, T. 1992, \apjl, 395, L83

\bibitem[{{Nathanail} \& {Contopoulos}(2015)}]{2015MNRAS.453L...1N}
{Nathanail}, A., \& {Contopoulos}, I. 2015, \mnras, 453, L1

\bibitem[{{Nava} {et~al.}(2011){Nava}, {Ghirlanda}, {Ghisellini}, \& {Celotti}}]{2011A&A...530A..21N}
{Nava}, L., {Ghirlanda}, G., {Ghisellini}, G., \& {Celotti}, A. 2011, \aap, 530, A21

\bibitem[{Nava {et~al.}(2013)Nava, Sironi, Ghisellini, Celotti, \& Ghirlanda}]{nava2013afterglow}
Nava, L., Sironi, L., Ghisellini, G., Celotti, A., \& Ghirlanda, G. 2013, Monthly Notices of the Royal Astronomical Society, 433, 2107, arXiv:1211.2806v2 [astro-ph.HE]

\bibitem[{{Newville} {et~al.}(2014){Newville}, {Stensitzki}, {Allen}, \& {Ingargiola}}]{2014zndo.....11813N}
{Newville}, M., {Stensitzki}, T., {Allen}, D.~B., \& {Ingargiola}, A. 2014, {LMFIT: Non-Linear Least-Square Minimization and Curve-Fitting for Python}, v0.8.0,  Zenodo, doi:10.5281/zenodo.11813

\bibitem[{{O'Connor} {et~al.}(2022){O'Connor}, {Troja}, {Dichiara}, {Beniamini}, {Cenko}, {Kouveliotou}, {Gonz{\'a}lez}, {Durbak}, {Gatkine}, {Kutyrev}, {Sakamoto}, {S{\'a}nchez-Ram{\'\i}rez}, \& {Veilleux}}]{2022MNRAS.515.4890O}
{O'Connor}, B., {Troja}, E., {Dichiara}, S., {et~al.} 2022, \mnras, 515, 4890

\bibitem[{{Oh} {et~al.}(2018){Oh}, {Koss}, {Markwardt}, {Schawinski}, {Baumgartner}, {Barthelmy}, {Cenko}, {Gehrels}, {Mushotzky}, {Petulante}, {Ricci}, {Lien}, \& {Trakhtenbrot}}]{2018ApJS..235....4O}
{Oh}, K., {Koss}, M., {Markwardt}, C.~B., {et~al.} 2018, \apjs, 235, 4

\bibitem[{{{\"O}zel} \& {Freire}(2016)}]{2016ARA&A..54..401O}
{{\"O}zel}, F., \& {Freire}, P. 2016, \araa, 54, 401

\bibitem[{{Perna} {et~al.}(2025){Perna}, {Gottlieb}, {Shukla}, \& {Radice}}]{2025PhRvD.111f3015P}
{Perna}, R., {Gottlieb}, O., {Shukla}, E., \& {Radice}, D. 2025, \prd, 111, 063015

\bibitem[{{Piran}(2004)}]{2004RvMP...76.1143P}
{Piran}, T. 2004, Reviews of Modern Physics, 76, 1143

\bibitem[{{Ravi} \& {Lasky}(2014)}]{2014MNRAS.441.2433R}
{Ravi}, V., \& {Lasky}, P.~D. 2014, \mnras, 441, 2433

\bibitem[{{Rezzolla} \& {Kumar}(2015)}]{2015ApJ...802...95R}
{Rezzolla}, L., \& {Kumar}, P. 2015, \apj, 802, 95

\bibitem[{{Romani} {et~al.}(2022){Romani}, {Kandel}, {Filippenko}, {Brink}, \& {Zheng}}]{2022ApJ...934L..17R}
{Romani}, R.~W., {Kandel}, D., {Filippenko}, A.~V., {Brink}, T.~G., \& {Zheng}, W. 2022, \apjl, 934, L17

\bibitem[{{Rowlinson} {et~al.}(2013){Rowlinson}, {O'Brien}, {Metzger}, {Tanvir}, \& {Levan}}]{rowlinson2013}
{Rowlinson}, A., {O'Brien}, P.~T., {Metzger}, B.~D., {Tanvir}, N.~R., \& {Levan}, A.~J. 2013, \mnras, 430, 1061

\bibitem[{{Rowlinson} {et~al.}(2010){Rowlinson}, {O'Brien}, {Tanvir}, {Zhang}, {Evans}, {Lyons}, {Levan}, {Willingale}, {Page}, {Onal}, {Burrows}, {Beardmore}, {Ukwatta}, {Berger}, {Hjorth}, {Fruchter}, {Tunnicliffe}, {Fox}, \& {Cucchiara}}]{2010MNRAS.409..531R}
{Rowlinson}, A., {O'Brien}, P.~T., {Tanvir}, N.~R., {et~al.} 2010, \mnras, 409, 531

\bibitem[{Sari {et~al.}(1998)Sari, Piran, \& Narayan}]{sari98}
Sari, R., Piran, T., \& Narayan, R. 1998, The Astrophysical Journal, 497, L17.
\newblock \url{https://ui.adsabs.harvard.edu/abs/1998ApJ...497L..17S}

\bibitem[{{Sarin} {et~al.}(2019){Sarin}, {Lasky}, \& {Ashton}}]{2019ApJ...872..114S}
{Sarin}, N., {Lasky}, P.~D., \& {Ashton}, G. 2019, \apj, 872, 114

\bibitem[{{Stergioulas} \& {Friedman}(1995)}]{1995ApJ...444..306S}
{Stergioulas}, N., \& {Friedman}, J.~L. 1995, \apj, 444, 306

\bibitem[{{Takami} {et~al.}(2015){Takami}, {Rezzolla}, \& {Baiotti}}]{2015PhRvD..91f4001T}
{Takami}, K., {Rezzolla}, L., \& {Baiotti}, L. 2015, \prd, 91, 064001

\bibitem[{{Tang} {et~al.}(2019){Tang}, {Huang}, {Geng}, \& {Zhang}}]{2019ApJS..245....1T}
{Tang}, C.-H., {Huang}, Y.-F., {Geng}, J.-J., \& {Zhang}, Z.-B. 2019, \apjs, 245, 1

\bibitem[{{Troja}(2019)}]{2019cxo..prop.5648T}
{Troja}, E. 2019, {The Collimation and Energetics of Short Grbs: Searching for Jet-Breaks with Chandra}, Chandra Proposal ID \#21500589, ,

\bibitem[{{Troja} {et~al.}(2007){Troja}, {Cusumano}, {O'Brien}, {Zhang}, {Sbarufatti}, {Mangano}, {Willingale}, {Chincarini}, {Osborne}, {Marshall}, {Burrows}, {Campana}, {Gehrels}, {Guidorzi}, {Krimm}, {La Parola}, {Liang}, {Mineo}, {Moretti}, {Page}, {Romano}, {Tagliaferri}, {Zhang}, {Page}, \& {Schady}}]{2007ApJ...665..599T}
{Troja}, E., {Cusumano}, G., {O'Brien}, P.~T., {et~al.} 2007, \apj, 665, 599

\bibitem[{{Troja} {et~al.}(2016){Troja}, {Sakamoto}, {Cenko}, {Lien}, {Gehrels}, {Castro-Tirado}, {Ricci}, {Capone}, {Toy}, {Kutyrev}, {Kawai}, {Cucchiara}, {Fruchter}, {Gorosabel}, {Jeong}, {Levan}, {Perley}, {Sanchez-Ramirez}, {Tanvir}, \& {Veilleux}}]{2016ApJ...827..102T}
{Troja}, E., {Sakamoto}, T., {Cenko}, S.~B., {et~al.} 2016, \apj, 827, 102

\bibitem[{{Tsang} {et~al.}(2012){Tsang}, {Read}, {Hinderer}, {Piro}, \& {Bondarescu}}]{2012PhRvL.108a1102T}
{Tsang}, D., {Read}, J.~S., {Hinderer}, T., {Piro}, A.~L., \& {Bondarescu}, R. 2012, \prl, 108, 011102

\bibitem[{{von Kienlin} {et~al.}(2020){von Kienlin}, {Meegan}, {Paciesas}, {Bhat}, {Bissaldi}, {Briggs}, {Burns}, {Cleveland}, {Gibby}, {Giles}, {Goldstein}, {Hamburg}, {Hui}, {Kocevski}, {Mailyan}, {Malacaria}, {Poolakkil}, {Preece}, {Roberts}, {Veres}, \& {Wilson-Hodge}}]{2020ApJ...893...46V}
{von Kienlin}, A., {Meegan}, C.~A., {Paciesas}, W.~S., {et~al.} 2020, \apj, 893, 46

\bibitem[{{Wang} {et~al.}(2024){Wang}, {Moradi}, \& {Liang}}]{2024ApJ...974...89W}
{Wang}, Y., {Moradi}, R., \& {Liang}, L. 2024, \apj, 974, 89

\bibitem[{{Xiao} \& {Dai}(2019)}]{2019ApJDai}
{Xiao}, D., \& {Dai}, Z.-G. 2019, \apj, 878, 62

\bibitem[{{Yorgancioglu} {et~al.}(2025){Yorgancioglu}, {Saeed}, {Moradi}, \& {Wang}}]{Yorgancioglu2025Dainotti}
{Yorgancioglu}, E.~S., {Saeed}, D.~M., {Moradi}, R., \& {Wang}, Y. 2025, arXiv e-prints, arXiv:2507.09292

\bibitem[{{Zhang}(2018)}]{2018pgrb.book.....Z}
{Zhang}, B. 2018, {The Physics of Gamma-Ray Bursts}, doi:10.1017/9781139226530

\bibitem[{{Zhang} \& {M{\'e}sz{\'a}ros}(2001)}]{2001ApJ...552L..35Z}
{Zhang}, B., \& {M{\'e}sz{\'a}ros}, P. 2001, \apjl, 552, L35

\bibitem[{{Zhang} {et~al.}(2018){Zhang}, {Lei}, {Zhang}, {Chen}, {Xiong}, \& {Song}}]{2018MNRAS.475..266Z}
{Zhang}, Q., {Lei}, W.~H., {Zhang}, B.~B., {et~al.} 2018, \mnras, 475, 266

\end{thebibliography}
\end{document}